\documentclass[fleqn,usenatbib,useAMS]{mnras}

\usepackage[T1]{fontenc}

\DeclareRobustCommand{\VAN}[3]{#2}
\let\VANthebibliography\thebibliography
\def\thebibliography{\DeclareRobustCommand{\VAN}[3]{##3}\VANthebibliography}

\usepackage{graphicx}	
\usepackage{amsmath}	

\usepackage{newtxtext,newtxmath}

\title[Crystallization-driven dynamos]{Slow convection and fast rotation in crystallization-driven white dwarf dynamos}

\author[S. Ginzburg et al.]{
Sivan Ginzburg,$^{1}$\thanks{E-mail: sivan.ginzburg@mail.huji.ac.il}\thanks{51 Pegasi b Fellow.} Jim Fuller$^{1}$, Adela Kawka$^{2}$ and Ilaria Caiazzo$^{1}$\thanks{Sherman Fairchild Fellow.}
\\
$^{1}$TAPIR, California Institute of Technology, Mailcode 350-17, Pasadena, CA 91125, USA\\
$^{2}$International Centre for Radio Astronomy Research, Curtin University, GPO Box U1987, Perth, WA 6845, Australia
}

\date{Accepted XXX. Received YYY; in original form ZZZ}

\pubyear{2021}

\begin{document}
\label{firstpage}
\pagerange{\pageref{firstpage}--\pageref{lastpage}}
\maketitle

\begin{abstract}
It has been recently suggested that white dwarfs generate magnetic fields in a process analogous to the Earth. The crystallization of the core creates a compositional inversion that drives convection, and combined with rotation, this can sustain a magnetic dynamo. We reanalyse the dynamo mechanism, arising from the slow crystallization of the core, and find convective turnover times $t_{\rm conv}$ of weeks to months -- longer by orders of magnitude than previously thought. With white dwarf spin periods $P\ll t_{\rm conv}$, crystallization-driven dynamos are almost always in the fast rotating regime, where the magnetic field $B$ is at least in equipartition with the convective motion and is possibly further enhanced by a factor of $B\propto (t_{\rm conv}/P)^{1/2}$, depending on the assumed dynamo scaling law. We track the growth of the crystallized core using \textsc{mesa} and compute the magnetic field $B(T_{\rm eff})$ as a function of the white dwarf's effective temperature $T_{\rm eff}$. We compare this prediction with observations and show that crystallization-driven dynamos can explain some -- but not all -- of the $\sim$MG magnetic fields measured for single white dwarfs, as well as the stronger fields measured for white dwarfs in cataclysmic variables, which were spun up by mass accretion to short $P$. Our $B(T_{\rm eff})$ curves might also explain the clustering of white dwarfs with Balmer emission lines around $T_{\rm eff}\approx 7500\textrm{ K}$.
\end{abstract}




\section{Introduction}\label{sec:intro}

Strong magnetic fields of $\sim \! 10^5-10^9$ G have been measured for a significant fraction of white dwarfs, including both single stars and cataclysmic variables (CVs), in which a white dwarf accretes mass from a companion star \citep[see][for reviews]{Ferrario2015,Ferrario2020}. 
The magnetic CVs are divided into `polars' -- in which the magnetic field synchronizes the white dwarf's spin with its orbit, and `intermediate polars' -- in which the white dwarf rotates much faster.
The origin of white dwarf magnetism is unclear. One possibility is that the observed white dwarf fields are fossils, reflecting the magnetic flux of their progenitor main sequence stars \citep{Angel81,BraithwaiteSpruit2004,Tout2004,WickramasingheFerrario2005}. Alternatively, magnetic fields could be generated during a common envelope phase
\citep[in which a binary companion is either destroyed or survives on a tight orbit; see][]{RegosTout95,Tout2008,Nordhaus2011}, or during the merger of two white dwarfs \citep{GarciaBerro2012}. All of these theories, however, have difficulty explaining the magnetic fields observed in post-common envelope binaries \citep{Schreiber2021Nat}.

\citet{Isern2017} proposed a different tantalizing scenario -- white dwarfs may generate magnetic fields similarly to the Earth's geodynamo \citep[e.g.][]{Stevenson83,ListerBuffett95,GlatzmaierRoberts97}. When white dwarfs cool down sufficiently, their interiors begin to crystallize (solidify), creating an unstable compositional gradient in the liquid above the crystallized core. This drives convective flows \citep{Stevenson1980,Mochkovitch83} which, combined with the white dwarf's rotation, can sustain a magnetic dynamo. A critical parameter in determining the dynamo's strength is the ratio of rotation to convection time-scales, given by the convective Rossby number ${\rm Ro}\equiv P/t_{\rm conv}$ ($P$ is the white dwarf's spin period and $t_{\rm conv}$ is the convective turnover time). \citet{Isern2017} found that for most white dwarfs $P\gg t_{\rm conv}$, limiting the dynamo field to $B\lesssim 10^5$ G. 

In a series of papers, \citet{Schreiber2021Nat,Schreiber2021} and \citet{Belloni2021} demonstrated that strong rotation-dependent dynamos operating during white dwarf crystallization may explain the occurrence rates of magnetism in different classes of single and accreting white dwarfs. Specifically, metal-polluted white dwarfs (which may have been spun up by planetary material accretion) are magnetic almost exclusively at effective temperatures $T_{\rm eff}\lesssim 8000$ K \citep{Hollands2015,Kawka2019}, coincident with the crystallization of their cores. In addition, a large fraction of CVs contain a strongly magnetized white dwarf, in stark contrast to their close but still detached binary progenitors \citep{Liebert2005,Liebert2015,Pala2020}. This observation poses a serious challenge to other theories, in which magnetic fields originate during the white dwarf's birth \citep{BelloniSchreiber2020}. 
In contrast, stable mass accretion from a companion spins up the white dwarf and may activate or enhance a magnetic dynamo, potentially explaining the prevalence of magnetism in CVs compared to their detached counterparts.

Despite these supporting arguments, \citet{Schreiber2021Nat,Schreiber2021} arbitrarily magnified the dynamo magnetic field, as calculated by \citet{Isern2017}, by orders of magnitude in order to match the measured fields of observed white dwarfs (single and accreting). Moreover, a critical ingredient of the theory, which is also required to match the observations, is a strong dependence of the crystallization-driven dynamo on rotation. However, such a dependence has not been calculated or tested quantitatively, but only qualitatively invoked.

Here, we reanalyse the crystallization-driven dynamo mechanism in white dwarfs and argue that \citet{Isern2017} have greatly underestimated the turnover time $t_{\rm conv}$. By consistently analysing convection arising from the slow crystallization of the white dwarf's core -- a process that unfolds over Gyrs -- we find that most white dwarfs are in fact in the fast rotating regime, with $P\ll t_{\rm conv}$ \citep[the opposite regime of][]{Isern2017}. The magnetic fields of such fast rotating dynamos may reach $B\sim 10^5-10^8$ G, depending on the white dwarf's rotation period $P$ and on the assumed dynamo scaling law. Thus, our work provides quantitative justification to the strong magnetic fields invoked by \citet{Schreiber2021Nat,Schreiber2021}, and their dependence on rotation.

The remainder of this paper is organized as follows. In Section \ref{sec:convection} we calculate the velocities and turnover times of the convective flows driven by the crystallization of a white dwarf's core. In Section \ref{sec:magnetic} we compute the magnetic field $B(P,T_{\rm eff})$ of the resulting dynamo, which we compare with observations in Section \ref{sec:obs}. We summarize our findings in Section \ref{sec:summary}.

\section{Crystallization driven convection}\label{sec:convection}

In this section we follow \citet{Isern2017}, as well as standard mixing-length theory, and discuss how the crystallization of the white dwarf's core drives convection. In Section \ref{sec:buoyancy} we improve the buoyancy calculation of \citet{Isern2017} and find turnover times that are longer by many orders of magnitude. We derive the same result using energy considerations in Section \ref{sec:flux}. We describe our numerical calculation and estimate typical velocities and turnover times in Section \ref{sec:typical}. The effect of rotation is discussed in Section \ref{sec:rotation}.

\subsection{Buoyancy}\label{sec:buoyancy}

When the core of a CO white dwarf crystallizes, the solid (crystal) phase is enriched in oxygen. This leaves behind a carbon-rich liquid that is lighter than the ambient CO mixture above -- driving a Rayleigh--Taylor instability \citep{Stevenson1980, Isern97, Isern2017}.

As the white dwarf cools down, the crystallized core grows; we mark its instantaneous outer radius and mass by $r$ and $m$. The size $l$ of the rising (Rayleigh--Taylor unstable) fluid elements above the core is limited by the core's size, by the scale height $h$, and by the width of the convection zone $\Delta r\equiv r_{\rm out}-r$, such that
\begin{equation}\label{eq:length}
    l\sim \min(r,h,\Delta r).
\end{equation}
At late times, i.e. low $T_{\rm eff}$, convection reaches the inner edge of the white dwarf's helium envelope. At earlier times, however, the outer edge of the convection zone $r_{\rm out}$ is smaller due to the increasing C to O ratio in the outer layers of the white dwarf \citep{Salaris97,Isern2017}. See Appendix \ref{sec:r_out} for how we calculate $r_{\rm out}$.

\citet{Isern2017} assumed that the density contrast $\Delta\rho/\rho$ of these rising elements compared to the ambient fluid is equal to the density difference between the carbon-rich liquid that is left behind after crystallization and the unperturbed CO mixture above, which is $\Delta\rho_0/\rho\sim 10^{-3}$ (\citealt{Isern2017}; see also \citealt{Mochkovitch83}). We, on the other hand, argue that the density contrast is much lower, because only a small fraction of a rising element crystallizes on a convective turnover time $t_{\rm conv}$, before the element mixes with the ambient fluid. In other words, as soon as an element begins to crystallize, the remaining slightly carbon-enriched liquid will begin to rise buoyantly, not allowing enough time for the element to reach the maximal contrast possible, $\Delta\rho_0/\rho$.

The actual density contrast is given by
\begin{equation}\label{eq:delta_rho}
    \frac{\Delta \rho}{\rho}=\frac{\Delta \rho_0}{\rho}\frac{\dot{m}t_{\rm conv}}{4\upi r^2l\rho} = \frac{\Delta \rho_0}{\rho} \frac{\dot{r}}{v_{\rm conv}}\,
\end{equation}
where $\dot{m}$ and $\dot{r}$ are the mass and radius growth rates of the crystallized core, $\rho$ is the density right above it, and
\begin{equation}
\label{eq:v_conv_def}
    v_{\rm conv} = \frac{l}{t_{\rm conv}}
\end{equation}
is the mixing length theory convective velocity. The rightmost term in equation \eqref{eq:delta_rho} indicates the fraction of the fluid element that crystallizes on a turnover time.

To be more accurate, the relevant mass of extra carbon remaining after crystallization is $m_{\rm rem}=m_{\rm liq}-m_{\rm sol}$, where $m_{\rm sol}= \dot{m}t_{\rm conv}$ is the solidified mass during one convective turnover time, and $m_{\rm liq}$ is the original liquid mass (from which oxygen has been depleted) before crystallization. From the conservation of oxygen, $x_{\rm liq}^{\rm O}m_{\rm liq}=x_{\rm sol}^{\rm O}m_{\rm sol}$, such that $m_{\rm rem}=m_{\rm sol}(x_{\rm sol}^{\rm O}/x_{\rm liq}^{\rm O}-1)\approx 0.3 m_{\rm sol}$, with the mass fractions of oxygen in the solid and liquid phases $x_{\rm sol/liq}^{\rm O}$ evaluated in Appendix \ref{sec:r_out}. We omit this mass correction factor of $\approx 0.3$ in equation \eqref{eq:delta_rho} for simplicity and due to other order-unity uncertainties of the mixing-length theory. We show below that our nominal magnetic field (Section \ref{sec:beyond}) scales as $B\propto v_{\rm conv}^{1/2}\propto \dot{m}^{1/6}$, such that the correction factor to $B$ is $\approx 0.3^{1/6}\approx 0.8$, which can be neglected given the other uncertainties of the model. 
Note that 
equation \eqref{eq:delta_rho}
can be also obtained using mass conservation, and assuming that convective upflows and downflows have similar filling factors (equation 1 of \citealt{MoffattLoper94}).

A lighter element rises upward with a buoyant acceleration $g\Delta\rho/\rho$, where $g$ is the local gravity, until it mixes and dissipates after traversing roughly its own length $l$, i.e. after interacting with its own mass of ambient fluid. The convective velocity it reaches is therefore
\begin{equation}\label{eq:v_conv}
    v_{\rm conv}\sim \left(\frac{\Delta\rho}{\rho}gl\right)^{1/2}=\left(g\frac{\Delta\rho_0}{\rho}\frac{\dot{m}t_{\rm conv}}{4\upi r^2\rho}\right)^{1/2}.
\end{equation}
We equate equations \eqref{eq:v_conv_def} and \eqref{eq:v_conv}
to find that
\begin{equation}\label{eq:tconv}
    t_{\rm conv}\sim\left[\frac{4\upi r^4l^2\rho}{Gm\dot{m}}\left(\frac{\Delta \rho_0}{\rho}\right)^{-1}\right]^{1/3},
\end{equation}
where we have substituted $g=Gmr^{-2}$ ($G$ is the gravitational constant), and with all the physical quantities evaluated at the edge of the crystallized core.

One interpretation of equation \eqref{eq:delta_rho} is crystallization of small oxygen-rich flakes inside a layer of width $l$ above the crystallized core, where the thermodynamic conditions are similar ($l\leq h$). The solid flakes settle down onto the core, leaving behind a homogeneous (well-mixed, because the flakes are small) slightly carbon-enriched liquid layer with a density deficit of $\Delta\rho$ \citep{Stevenson1980,Mochkovitch83}. In an alternative picture, crystallization occurs directly at the border of the solid core, leaving behind a very thin shell of concentrated carbon-rich liquid above it, with a much larger density contrast $\Delta\rho_0$ \citep{Loper78}. The material in this shell will rise buoyantly at an acceleration $g\Delta\rho_0/\rho$, as assumed by \citet{Isern2017}. However, the rising fluid elements from such a thin shell will be comparable in size to its width $w$, and will therefore quickly mix after rising a similar distance $w\ll l$. Such fast mixing continues until the largest convecting elements (of size $l$) homogenize with a $\Delta\rho\ll\Delta\rho_0$. In the limit of a very thin shell (as in the limit of very small flakes), the density contrast of the dominant length-scale for convection $l$ is controlled by the rate of crystallization, as assumed in equation \eqref{eq:delta_rho}, rather than by $w/l$. 

Further motivation for equation \eqref{eq:delta_rho} can be found by application to the Earth's convective liquid outer core, which lies on top of a solid inner core with a radius $r\approx 1.2\times 10^8\textrm{ cm}$ \citep{Engdahl74}. By applying the equations in section 2 of \citet{Isern2017} to the Earth's core, we find $v_{\rm conv}^{\rm theory}\approx\sqrt{0.2(3/8)g(\Delta\rho_0/\rho)(0.1 r)}\approx 3\times 10^3\textrm{ cm s}^{-1}$, with $\Delta\rho_0/\rho\approx 0.05$ from \citet{MoffattLoper94}. This velocity is orders of magnitude higher than the $v_{\rm conv}^{\rm actual}\sim 10^{-1}\textrm{ cm s}^{-1}$ inferred from measurements \citep[see, e.g.][and references therein]{FinlayAmit2011,Schaeffer2017}.
By using equations \eqref{eq:delta_rho} and \eqref{eq:v_conv}, we calculate a velocity reduction factor of
$(\Delta\rho/\Delta\rho_0)^{1/2}\sim(\dot{r}/v_{\rm conv}^{\rm actual})^{1/2}\sim 10^{-4}$, with $\dot{r}\sim 10^{-9}\textrm{ cm s}^{-1}$ 
assuming the Earth's core solidified over a few Gyrs. This correction factor brings theory and measurements into agreement, further justifying our equation \eqref{eq:delta_rho}.

None the less, crystallization-driven convection remains a complex problem both in the geophysical context and in white dwarfs, and it is not clear whether the same scaling relations apply in both cases. The detailed structure of the crystallization front depends on the thermodynamic and compositional properties of the core \citep{Loper78}, which differ between the Earth and white dwarfs. A more rigorous treatment of crystallization-driven convection, and in fact of convection in general \citep[e.g.][]{M3b2016, KupkaMuthsam2017}, is beyond the scope of this work.

\subsection{Energy flux}\label{sec:flux}

An alternative method to derive the convective velocity is by considering the gravitational energy released by settling matter as the core crystallizes. The energy release rate is given by
\begin{equation}\label{eq:lgrav}
    L_{\rm grav}=\frac{Gm\dot{m}}{r}\frac{\Delta \rho_0}{\rho}\frac{l}{r}.
\end{equation}
This energy is carried away by the convective flux:
\begin{equation}\label{eq:flux}
    F_{\rm grav}=\frac{L_{\rm grav}}{4\upi r^2}=\frac{Gm\dot{m}}{4\upi r^3}\frac{\Delta \rho_0}{\rho}\frac{l}{r}=\rho v_{\rm conv}^3.
\end{equation}
From equation \eqref{eq:flux}, the convective velocity is
\begin{equation}\label{eq:vconv_full}
    v_{\rm conv}=\left(\frac{Gm\dot{m}l}{4\upi r^4\rho}\frac{\Delta \rho_0}{\rho}\right)^{1/3},
\end{equation}
which is identical to equations \eqref{eq:v_conv_def} and \eqref{eq:tconv}.

We emphasize that equation \eqref{eq:flux} is meant only as an analogy with standard mixing length theory, where a heat flux $F$ drives convective velocities of $v_{\rm conv} \sim (F/\rho)^{1/3}$. In our case, convection is driven by compositional inversion, and the energy flux can be conducted away. In other words, $v_{\rm conv}$ does not have to carry all the energy released during crystallization. In fact, crystallization releases energy in other forms, such as latent heat and change in chemical potentials, which are comparable to $L_{\rm grav}$ \citep{Isern97, Isern2000}. Specifically, see equation (19) of \citet{Isern97}, which is similar to our equation \eqref{eq:lgrav} when $l\sim r\sim r_{\rm wd}$ (see Section \ref{sec:typical}). The fact that the convective velocities are similar to what is needed to carry $L_{\rm grav}$ (rather than the total luminosity) is only presented to build intuition, with a more physical derivation of $v_{\rm conv}$ given in Section \ref{sec:buoyancy}.

\subsection{Numerical calculation and typical values}\label{sec:typical}

\begin{figure}
\includegraphics[width=\columnwidth]{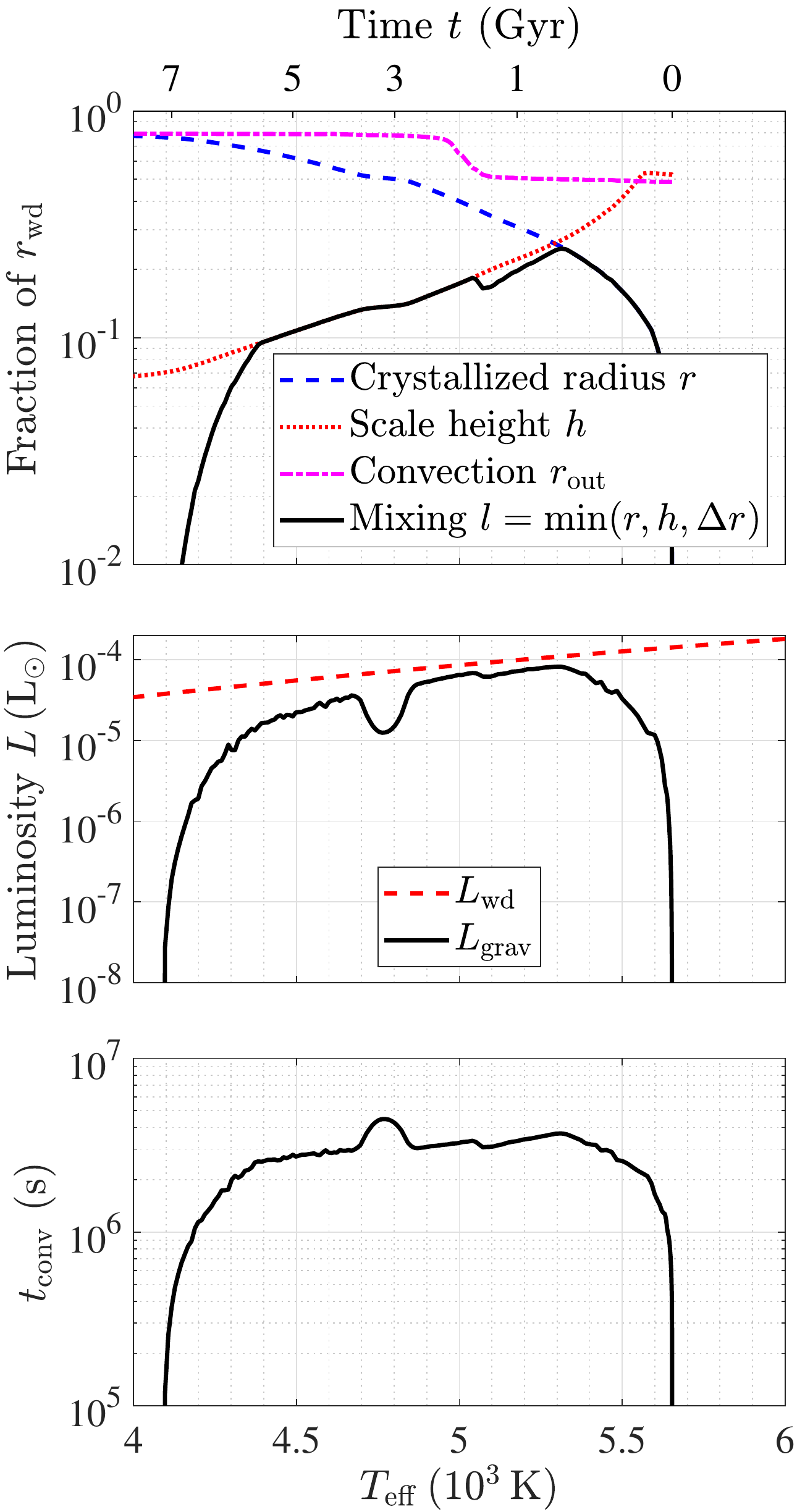}
\caption{Convection in a $0.6\,{\rm M}_\odot$ white dwarf as a function of its effective temperature $T_{\rm eff}$ and of the time elapsed since the onset of crystallization $t$. Top panel: length scales, given as a fraction of the white dwarf's radius $r_{\rm wd}$. The pressure scale height $h$ is evaluated at the edge of the crystallized core, which has a radius $r$. The typical mixing length for convection $l$ is given by the minimum of $r$, $h$, and the distance $\Delta r\equiv r_{\rm out}-r$ from $r$ to the outer edge of the convection zone $r_{\rm out}$. 
Middle panel: the gravitational settling power $L_{\rm grav}$ released by crystallization, calculated using equation \eqref{eq:lgrav}, compared to the white dwarf's total luminosity $L_{\rm wd}$. $L_{\rm grav}$ is roughly equal to the energy carried away by the convective motion. Bottom panel: the convective turnover time $t_{\rm conv}$ (equation \ref{eq:tconv}).}
\label{fig:scales}
\end{figure}

In Fig. \ref{fig:scales} we plot the relevant length scales for a crystallizing $0.6\,{\rm M}_\odot$ white dwarf, calculated using the stellar evolution code \textsc{mesa} \citep{Paxton2011,Paxton2013,Paxton2015,Paxton2018,Paxton2019}. Crystallization is assumed to occur when the plasma coupling parameter (the ratio of the Coulomb to thermal energy) exceeds $\Gamma=230$, consistent with the updated `Skye' equation of state, and appropriate for a CO mixture \citep{Bauer2020,Jermyn2021}. This value is somewhat higher than the traditional one-component $\Gamma = 175$ \citep{PotekhinChabrier2000,PotekhinChabrier2010}, resulting in lower crystallization temperatures. 
The mixing length $l$ initially follows the growing core's size $r$, until it becomes comparable (to within an order-unity factor) to the white dwarf's radius $r_{\rm wd}$. The mixing length then largely follows the gradually decreasing scale height $h$ at the core's edge, until the CO crystallization front approaches the inner edge of the helium layer, at which point $l$ sharply drops. At early times, the convection zone does not reach the helium layer (see Appendix \ref{sec:r_out}), slightly restricting $l\leq \Delta r$ near its peak. 
The slower crystallization $\dot{r}$ at $T_{\rm eff}\approx 4800\textrm{ K}$ and the resulting bumps in $L_{\rm grav}$ and $t_{\rm conv}$ correspond to a similar bump in the cooling delay due to latent heat release in `Skye' \citep[fig. 12 of][]{Jermyn2021}. This feature does not significantly affect our magnetic field estimate (Section \ref{sec:magfield}).

We note that \textsc{mesa} and `Skye' currently do not implement phase separation upon crystallization and only include the latent heat release \citep{Jermyn2021}. We estimate the convective velocity and the resulting magnetic field using buoyancy arguments (equation \ref{eq:v_conv}) in a post-processing procedure (the energy considerations in Section \ref{sec:flux} are given only for intuition). We are therefore sensitive only to the white dwarf's overall cooling rate through the growth of the crystallized core $\dot{m}$. The energy associated with phase separation is estimated at about 10 to 20 per cent of the total energy released during crystallization \citep{Isern2000}, such that we do not expect a large deviation from our estimates (from equation \ref{eq:vconv_full}, $v_{\rm conv}\propto\dot{m}^{1/3}$). This should be verified, however, by future work that includes phase separation explicitly in the cooling curve \citep[e.g.][]{Salaris2010}.

For a rough order of magnitude estimate of the typical velocities and time-scales, we use the approximation $l\sim r\sim h\sim\Delta r\sim r_{\rm wd}$, which breaks down at both very early and very late times. We also approximate $\dot{m}\sim m_{\rm wd}/t_{\rm cryst}$, where $m_{\rm wd}$ is the white dwarf's mass and $t_{\rm cryst}\sim{\rm Gyr}$ is the time to crystallize a significant portion of it (see Fig. \ref{fig:scales}). With these approximations, and omitting order unity coefficients, equation \eqref{eq:tconv} reads
\begin{equation}
    t_{\rm conv}\sim\left[t_{\rm dyn}^2t_{\rm cryst}\left(\frac{\Delta\rho_0}{\rho}\right)^{-1}\right]^{1/3}\sim 10^7\,{\rm s}, 
\end{equation}
where
\begin{equation}
    t_{\rm dyn}\equiv \left(\frac{r_{\rm wd}^3}{Gm_{\rm wd}}\right)^{1/2}\sim\left(\frac{1}{G\rho}\right)^{1/2} \sim 3\,{\rm s}
\end{equation}
is the white dwarf's dynamical time-scale (see Fig. \ref{fig:scales} for a more accurate computation of $t_{\rm conv}$). Similarly, equation \eqref{eq:vconv_full} simplifies to
\begin{equation}
    v_{\rm conv}\sim\left(G\dot{m}\frac{\Delta\rho_0}{\rho}\right)^{1/3}\sim\frac{r_{\rm wd}}{t_{\rm conv}}\sim 10^2\textrm{ cm s}^{-1}.
\end{equation}
The velocities that we find are lower by orders of magnitude compared to \citet{Isern2017} because, according to equation \eqref{eq:delta_rho}, the effective density contrast is reduced by a factor of $\Delta\rho/\Delta\rho_0\sim t_{\rm conv}/t_{\rm cryst}\ll 1$.

\subsection{Rotation}\label{sec:rotation}

The long turnover times $t_{\rm conv}$ estimated above imply that most observed white dwarfs are in the fast rotating regime, with spin periods $P\ll t_{\rm conv}$. In this regime, the Coriolis force constrains the convective motion. Specifically, \citet{Stevenson1979} and \citet{Barker2014} suggested that rotation reduces the convective velocity by a factor of
\begin{equation}\label{eq:rotation}
    \frac{v_{\rm conv}^{\rm rot}}{v_{\rm conv}}\sim\left(\frac{P}{t_{\rm conv}}\right)^{1/5},
\end{equation}
where $v_{\rm conv}$ and $t_{\rm conv}$ are evaluated without rotation, as before. Using this scaling for $v_{\rm conv}$ implies even longer convective turnover times, and could affect the scaling of the dynamo-generated field (Section \ref{sec:magfield}).

\section{Magnetic field}\label{sec:magnetic}
\label{sec:magfield}

\citet{Isern2017} suggested that the convective motion, driven by crystallization, inside a rotating white dwarf can power a magnetic dynamo, similarly to the Earth and other planets \citep[e.g.][]{MoffattLoper94,ListerBuffett95}. In Sections \ref{sec:equi} and \ref{sec:beyond} we calculate the magnetic field strength $B$ using two different dynamo scaling laws, and in Section \ref{sec:diffusion} we estimate the magnetic diffusion time to the white dwarf's surface.

\subsection{Equipartition}\label{sec:equi}

\citet{Christensen2009} suggested that in rapidly spinning stars and planets, with rotation periods $P\ll t_{\rm conv}$, the magnetic energy density reaches an equipartition with the kinetic energy in convective eddies
\begin{equation}\label{eq:equi}
    \frac{B^2}{8\upi}\sim \frac{1}{2}\rho v_{\rm conv}^2= \frac{1}{2}\rho^{1/3}F_{\rm grav}^{2/3},
\end{equation}
where the last equality is given by equation \eqref{eq:flux}.
For consistency with the \citet{Christensen2009} scaling, which is independent of rotation, we use $v_{\rm conv}$ rather than $v_{\rm conv}^{\rm rot}$ in equation \eqref{eq:equi}; the effects of rotation are discussed in Section \ref{sec:beyond}.
Using the typical velocities in Section \ref{sec:typical}, we estimate a typical $B\sim 10^5\,{\rm G}$. 

\begin{figure}
\includegraphics[width=\columnwidth]{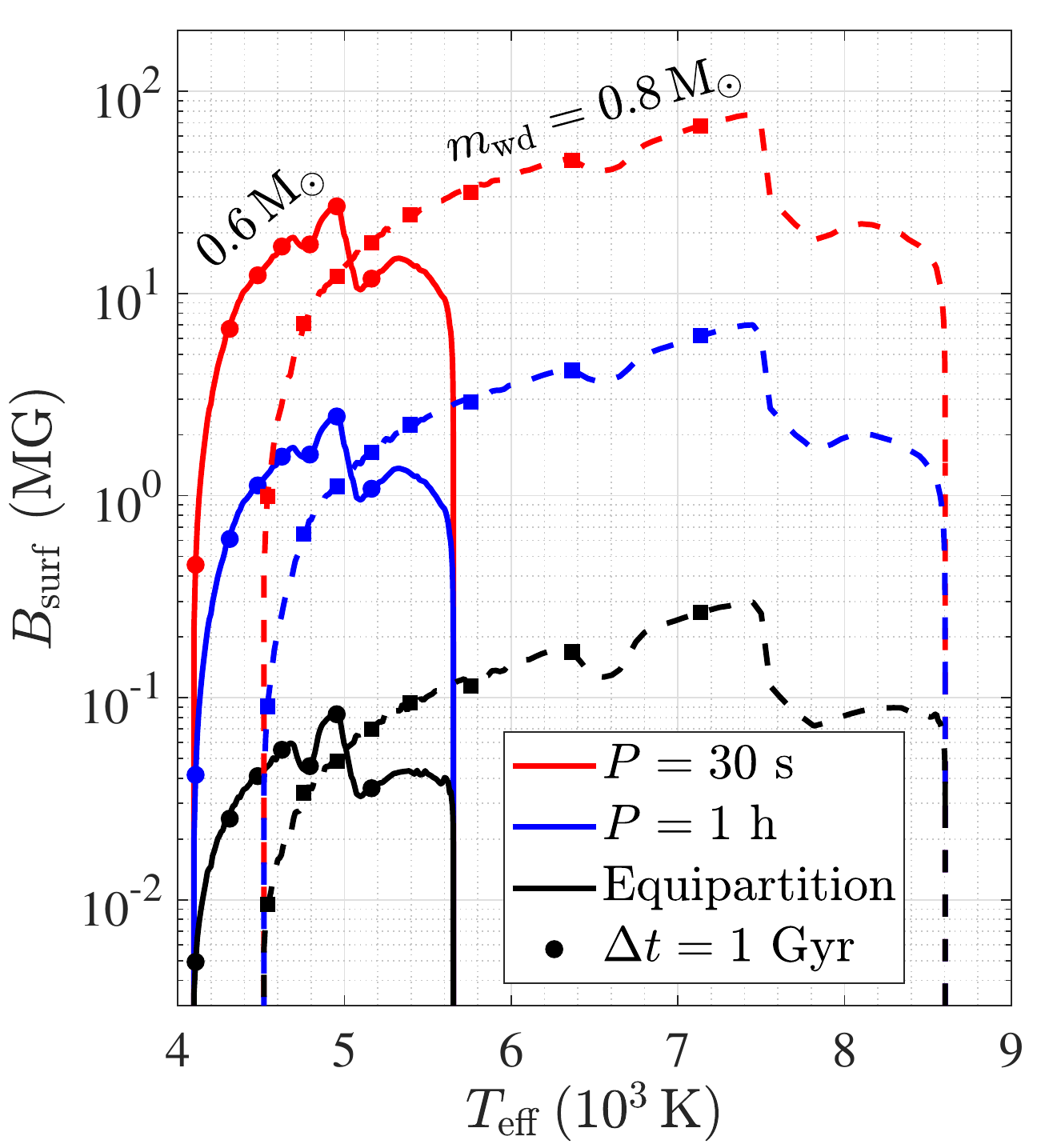}
\caption{Surface magnetic fields $B_{\rm surf}$ generated by crystallization driven dynamos as a function of the white dwarf's effective temperature $T_{\rm eff}$ for 0.6 (solid lines) and 0.8 ${\rm M}_\odot$ (dashed lines) white dwarfs. The surface and dynamo magnetic fields are related through equation \eqref{eq:B_surf}. The black (bottom) lines are plotted using equation \eqref{eq:equi}, which assumes equipartition between the magnetic energy and kinetic energy in convective eddies. The magnetic field in this case does not depend on rotation. The blue (middle) and red (top) lines are plotted using equation \eqref{eq:super_eq}, which assumes a balance between the Lorentz and Coriolis forces. This scaling may be more appropriate for the short white dwarf rotation periods $P\ll t_{\rm conv}$ that we consider. The markers (circles and squares) indicate 1 Gyr time intervals since the onset of crystallization.}
\label{fig:magnetic}
\end{figure}

In Fig. \ref{fig:magnetic} we plot $B_{\rm surf}(T_{\rm eff})$ which we calculated with \textsc{mesa} using equations \eqref{eq:vconv_full} and \eqref{eq:equi}. Specifically, $\dot{m}$ is given by an instantaneous numerical derivative of the growing crystallized core's mass $m$, and the mixing length $l$ is given by equation \eqref{eq:length}. 
The magnetic field at the surface $B_{\rm surf}$ is related to the dynamo field $B$ by
\begin{equation}\label{eq:B_surf}
    \frac{B_{\rm surf}}{B}=\left(\frac{r_{\rm out}}{r_{\rm wd}}\right)^3,
\end{equation}
where we have assumed a dipole geometry from the dynamo's top $r_{\rm out}$ to the surface $r_{\rm wd}$ \citep[similarly to][]{ReinersChristensen2010}. The curves are characterized by a steep rise to $\sim 0.1\,{\rm MG}$ at the onset of crystallization, followed by a further increase in $B_{\rm surf}$ when convection reaches the helium layer (see Fig. \ref{fig:scales}). The magnetic field then gradually declines as the white dwarf cools to lower $T_{\rm eff}$, until nearly all the CO crystallizes and the dynamo shuts off.

The hydrogen and helium layers of the white dwarfs in Fig. \ref{fig:magnetic} weigh $m_{\rm H}=3.7\times 10^{-5}\,{\rm M}_\odot$, $m_{\rm He}=2.2\times 10^{-2}\,{\rm M}_\odot$ for the 0.6 ${\rm M}_\odot$ white dwarf and  $m_{\rm H}=1.4\times 10^{-6}\,{\rm M}_\odot$, $m_{\rm He}=3.4\times 10^{-3}\,{\rm M}_\odot$ for the 0.8 ${\rm M}_\odot$ white dwarf. \citet{Isern2017} found that magnetic dynamos in hydrogen-deficient white dwarfs generate slightly stronger maximal fields due to differences in the cooling rate.

\citet{Isern2017} used the \citet{Christensen2009} scaling to relate the white dwarf's magnetic field to its convective flux, also finding $B\sim 0.1\,{\rm MG}$. However, their underestimated short convective turnover times $\sim \! 10^2$ s imply that only the fastest spinning white dwarfs (with $P$ shorter than about a minute) satisfy the condition $P\ll t_{\rm conv}$, which is required to reach equipartition according to \cite{Christensen2009}. All other white dwarfs are expected to generate weaker fields. \citet{Schreiber2021Nat} used this to explain the higher incidence of strong magnetic fields in CVs: only those would spin fast enough to be in the rapidly rotating limit such that the \citet{Christensen2009} scaling would apply. \citet{Schreiber2021Nat} then speculated that the unknown dependence of the dynamo on the magnetic Prandtl number could amplify $B$ a couple orders of magnitude beyond the \citet{Christensen2009} scaling -- explaining observed $\sim \! 10-100$ MG fields for white dwarfs in accreting systems.

In our consistent calculation of the convection, on the other hand, we find much longer $t_{\rm conv}\sim 10^6-10^7$ s (Section \ref{sec:convection}), such that most observed magnetic white dwarfs satisfy $P\ll t_{\rm conv}$ -- crystallization driven dynamos are naturally in the fast rotating regime, even for white dwarfs with normal spin periods. We therefore conclude that equipartition and $\sim \! 0.1$ MG fields are reached naturally by these dynamos (the black/bottom lines in Fig. \ref{fig:magnetic}), without resorting to an unknown Prandtl number dependence. In Section \ref{sec:beyond} we argue that our model can yield even stronger fields for fast rotators; specifically, we naturally reproduce $\sim \! 10 - 100$ MG fields in systems that have been spun up by accretion.

\subsection{Beyond equipartition}\label{sec:beyond}

Recent three-dimensional simulations suggest that rapidly rotating convective dynamos, with Rossby numbers ${\rm Ro}\equiv P/t_{\rm conv}\ll 1$, can reach super-equipartition magnetic fields \citep{Augustson2016,Augustson2019}. Specifically,
\begin{equation}\label{eq:super_eq}
\frac{B^2}{4\upi\rho v_{\rm conv}^2}\sim {\rm Ro}^{-1}=\frac{t_{\rm conv}}{P},    
\end{equation}
which can be intuitively understood by a balance between the Lorentz and Coriolis forces (equipartition can be understood as a balance between the Lorentz and inertial forces, but at ${\rm Ro}\ll 1$, the Coriolis force dominates over inertial forces).
Accounting for how the Coriolis force restricts rotating convective flows (Section \ref{sec:rotation}) would reduce $B\propto t_{\rm conv}^{1/2}v_{\rm conv}\propto v_{\rm conv}^{1/2}$ by a factor of $(v_{\rm conv}^{\rm rot}/v_{\rm conv})^{1/2}\sim(P/t_{\rm conv})^{1/10}$ according to equation \eqref{eq:rotation}. This would imply $B\propto P^{-2/5}$ instead of $B\propto P^{-1/2}$. It is not clear, however, whether magneto-hydrodynamic flows in which the Coriolis force is balanced by the Lorentz force are also restricted \citep{Christensen2010}. Since in any case the difference in $B$ is only a factor of a few at most, and the scaling with $P$ hardly changes, we keep equation \eqref{eq:super_eq} and treat it as an upper limit.

In Fig. \ref{fig:magnetic} we plot $B_{\rm surf}(T_{\rm eff})$ using equation \eqref{eq:super_eq}, with $t_{\rm conv}$ evaluated by equation \eqref{eq:tconv}. Rapid rotation magnifies the magnetic field beyond equipartition, reaching $B_{\rm surf}\sim 1-10$ MG for spin periods of hours, and up to $B_{\rm surf}\sim 10^2$ MG for faster rotation. This could explain the strong fields measured for white dwarfs in CVs \citep{Ferrario2015}, which may have been spun up by mass accretion to periods of $\sim 30$ s \citep[][]{Schreiber2021Nat}.

We emphasize that our model implies the opposite Rossby number regime of \citet{Isern2017}. Our long convective turnover times ensure ${\rm Ro}\ll 1$, whereas \citet{Isern2017} greatly underestimate $t_{\rm conv}$, leading to ${\rm Ro}\gtrsim 1$ for most systems. The consequences are stronger $B$ fields and a stronger dependence on the white dwarf's rotation period in our case $B\propto P^{-1/2}$, assuming equation \eqref{eq:super_eq} applies. While there is some dependence on $P$ in the ${\rm Ro}\gtrsim 1$ regime as well, it is considerably weaker \citep[see fig. 1 of][]{Augustson2016}. Recently, \citet{Brun2022} found a steeper dependence of about $B_{\rm surf}\propto {\rm Ro}^{-1.3}$ for the surface field at the top of their simulated convective dynamo compared to the volume-averaged bulk field $B\propto {\rm Ro}^{-0.5}$. This could imply stronger surface fields for rapidly rotating white dwarfs, with an even stronger dependence on rotation than our nominal Fig. \ref{fig:magnetic}.

\subsection{Magnetic diffusion}\label{sec:diffusion}

\begin{figure}
\includegraphics[width=\columnwidth]{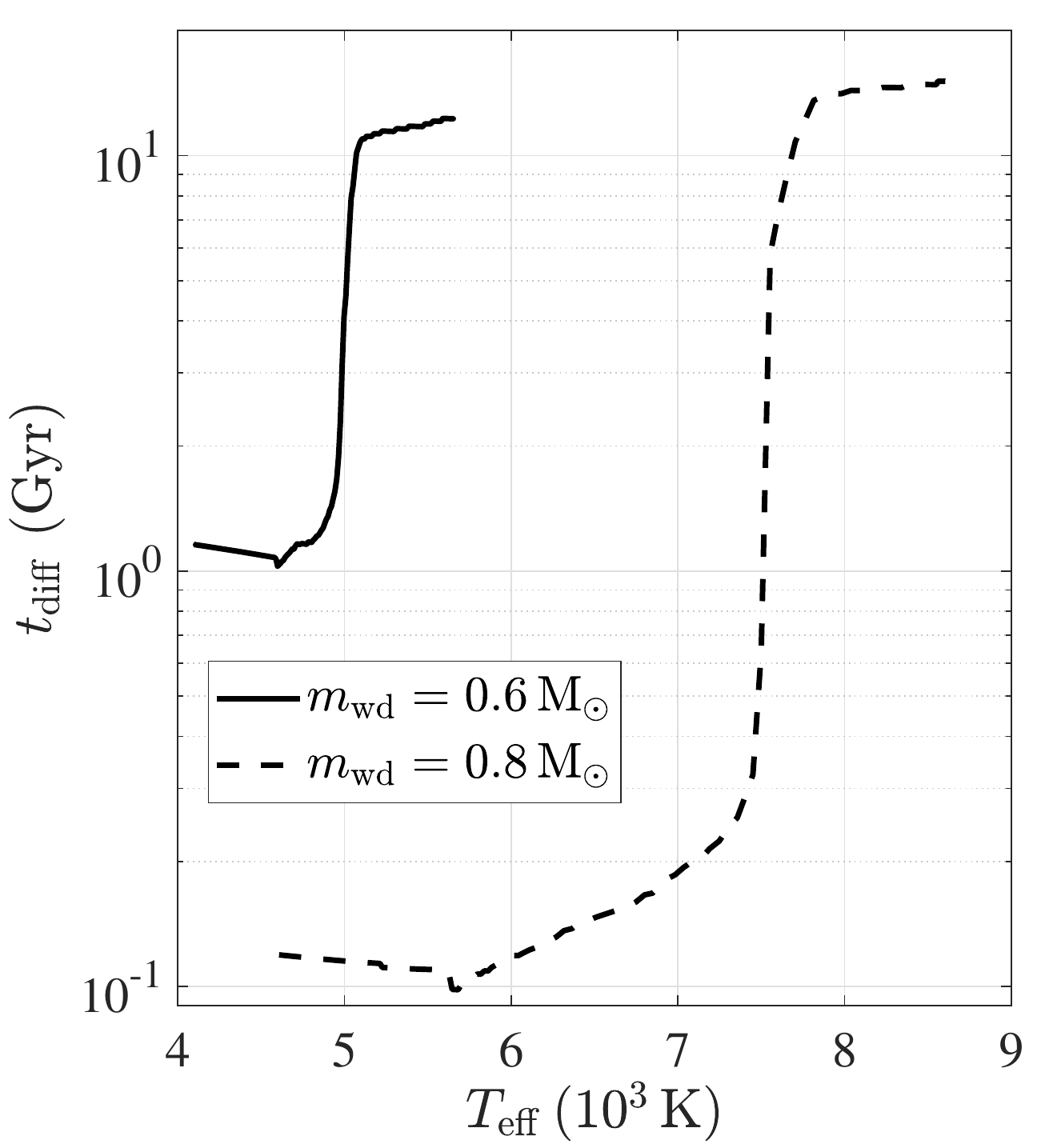}
\caption{The magnetic diffusion time $t_{\rm diff}$ from the outer edge of the convection zone $r_{\rm out}$ to the white dwarf's surface, calculated using equation \eqref{eq:tdiff} for  
$0.6$ (solid line) and $0.8\,{\rm M}_\odot$ (dashed line) white dwarfs during the operation of crystallization driven dynamos. The sharp drop at $T_{\rm eff}\approx 5000\textrm{ K}$ or $T_{\rm eff}\approx 7500\textrm{ K}$ (depending on $m_{\rm wd}$) is caused by the expansion of convection to the edge of the helium envelope (see Appendix \ref{sec:r_out}).}
\label{fig:diffusion}
\end{figure}

As explained in Appendix \ref{sec:r_out}, steep compositional gradients limit the convection to a radius $r_{\rm out}<r_{\rm wd}$ (see Fig. \ref{fig:scales}). Can the dynamo-generated magnetic field penetrate from there to the surface, where it can be observed? The finite Ohmic resistivity enables magnetic fields to slowly diffuse outwards. 

In Fig. \ref{fig:diffusion} we estimate the magnetic diffusion time-scale from $r_{\rm out}$ to the surface
\begin{equation}\label{eq:tdiff}
    t_{\rm diff}\sim\int_{r_{\rm out}}^{r_{\rm wd}}{\frac{{\rm d}(r-r_{\rm out})^2}{\eta(r)}=\int_{r_{\rm out}}^{r_{\rm wd}}\frac{2(r-r_{\rm out}){\rm d}r}{\eta(r)}}.
\end{equation}
The magnetic diffusivity $\eta$ is computed similarly to \citet{Cantiello2016}, by interpolating between expressions that are valid in the non-degenerate, partially degenerate, and fully degenerate regimes \citep{Spitzer62,NandkumarPethick84,Wendell1987}. 
Fig. \ref{fig:diffusion} shows that $t_{\rm diff}$ is initially very long. However, once the convection zone reaches the edge of the helium envelope, $t_{\rm diff}$ drops and becomes shorter than the dynamo's operation time (see Fig. \ref{fig:magnetic}), allowing the magnetic field to emerge at the surface. 

At the same time, $t_{\rm diff}\gg t_{\rm conv}$ \citep[with even longer diffusion times deeper in the convective interior; e.g.][]{Cumming2002}, such that crystallization-driven convection is in the high magnetic Reynolds number regime ($\sim\!10^{11}$), as assumed by our dynamo scaling laws. The Earth's magnetic Reynolds number is $\sim\!10^{3}$ \citep{Christensen2010,Davies2015}, much closer to the critical value of $\sim\! 50$ \citep[see][]{ChristensenAubert2006,Christensen2010}, possibly indicating that the geodynamo follows a somewhat different scaling.

Other processes, such as advection \citep{CharbonneauMacGregor2001} or magnetic buoyancy \citep{MacGregorCassinelli2003,MacDonaldMullan2004} may transport magnetic fields through the non-convecting layers on shorter time-scales. These mechanisms have been invoked in the context of massive stars, in which magnetic fields generated by a dynamo in the convective core have to penetrate through the radiative envelope to be observed.   
\section{Comparison with observations}\label{sec:obs}

\subsection{Single white dwarfs}

\begin{figure}
\includegraphics[width=\columnwidth]{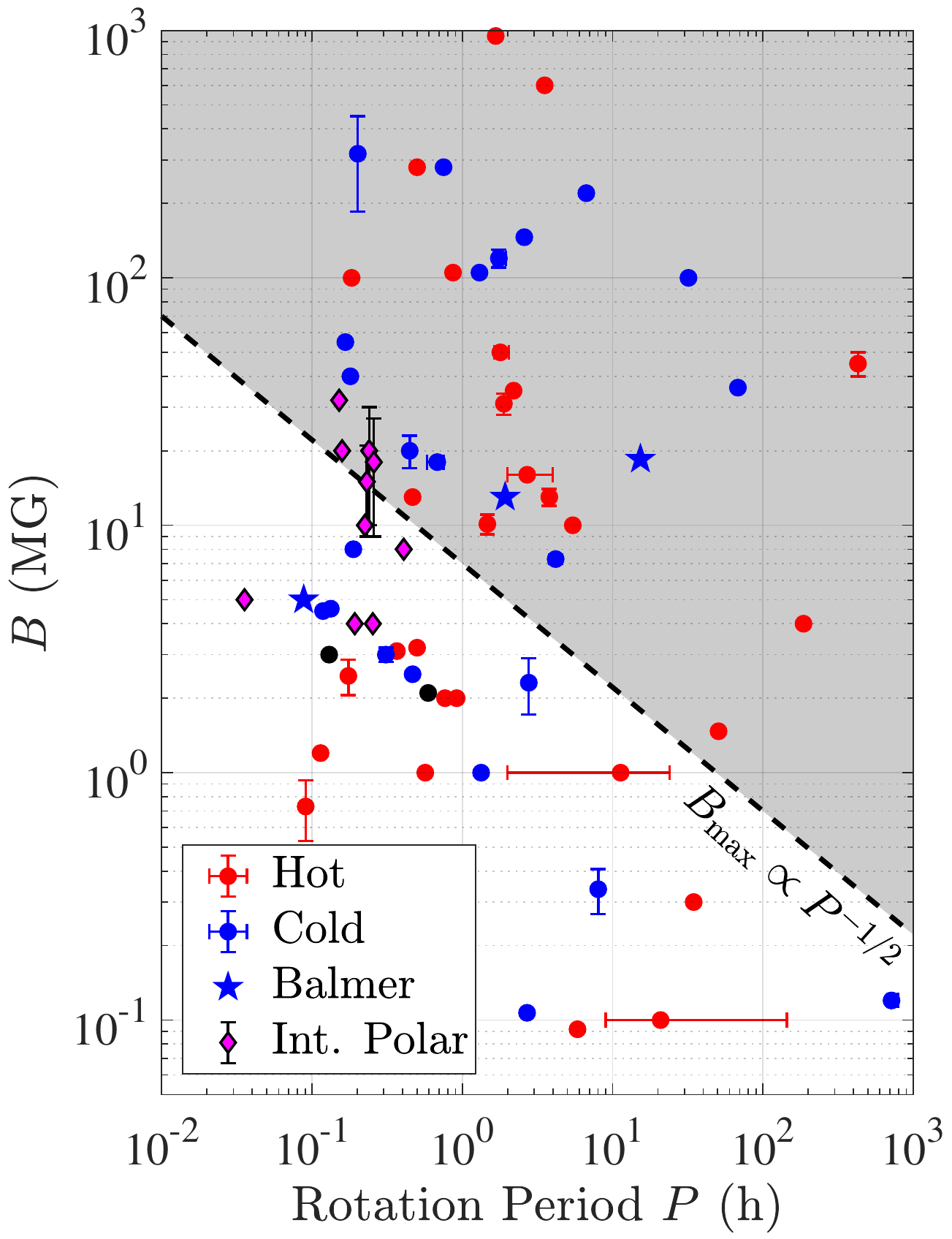}
\caption{Magnetic white dwarfs with measured rotation periods $P$. Single white dwarfs are from \citet[][excluding extremely slow rotators with inferred $P>10^5$ h]{Kawka2020} and Caiazzo et al. (in preparation). They are divided into stars that are cold enough (for their mass) to have started crystallizing (blue circles) and stars that are too hot (red circles). The critical $T_{\rm eff}(m_{\rm wd})$ for crystallization is given by \citet{Schreiber2021Nat}, which is consistent with our Fig. \ref{fig:magnetic}. The black circles indicate white dwarfs without a measured mass or temperature.
The blue star markers indicate single white dwarfs that exhibit Zeeman-split Balmer emission lines \citep{Gansicke2020,Walters2021}. Intermediate polars (magenta diamonds) are from \citet{Ferrario2015}. We do not include polars because they have been spun down after the generation of the magnetic field. 
The dashed black line is given by equation \eqref{eq:super_eq}, normalized to the maximal surface magnetic field of a $0.8\,\rm{M}_\odot$ crystallizing white dwarf, as calculated in Fig. \ref{fig:magnetic}. The crystallization-driven dynamo theory may account for the magnetic fields of cold white dwarfs below the line, i.e. blue and potentially magenta markers in the unshaded region. Their measured magnetic fields are anti-correlated with their periods, as predicted by the theory.}
\label{fig:kawka}
\end{figure}

In Fig. \ref{fig:kawka} we compare our rotationally enhanced (i.e. ${\rm Ro}\ll1$) fields to the population of magnetic white dwarfs with measured spin periods. Evidently, many of the measured magnetic fields cannot be explained by crystallization-driven dynamos: they are either well above our maximum $B_{\rm max}$, or the white dwarfs are too hot to be crystallizing. We conclude that the magnetic fields of at least some white dwarfs are produced by other mechanisms (Section \ref{sec:intro}). None the less, crystallization-driven dynamos may account for a significant fraction of the sample -- blue markers below and possibly right above the dashed black line. The magnetic fields of these white dwarfs seem to decrease with the rotation period $P$, as predicted by equation \eqref{eq:super_eq}, though this is of course partially by definition. We caution that even the sub-sample that is consistent with a crystallization-driven dynamo may in principle be contaminated by other sources of magnetism. Also, there might be an observational bias against rapidly rotating white dwarfs with weak magnetic fields because of rotational line broadening.    

In addition to the white dwarfs presented in Fig. \ref{fig:kawka}, we can also reproduce the $B\sim 1-10$ MG fields measured for metal-polluted white dwarfs, which were plausibly spun up to periods of minutes to hours by planetary material accretion \citep{Schreiber2021}. Almost all of these metal-polluted magnetic white dwarfs have $T_{\rm eff}< 8000\textrm{ K}$, strongly supporting a crystallization-driven dynamo scenario. 
Unlike \citet{Schreiber2021Nat,Schreiber2021}, we do not have to postulate a magnetic field enhancement due to the white dwarf's Prandtl number to reproduce the observations -- instead, our longer $t_{\rm conv}$ naturally leads to a rotational enhancement. In fact, white dwarf magnetic Prandtl numbers are of order unity \citep{Isern2017}, and are therefore in the same regime as the simulations examined by \citet{Augustson2019}, for which equation \eqref{eq:super_eq} applies. Moreover, \citet{Augustson2019} do not find a significant dependence on the Prandtl number in this regime.  
 
Recently, a small class of magnetic white dwarfs that exhibit Zeeman-split Balmer emission lines has been identified \citep{GreensteinMcCarthy85,Gansicke2020,Reding2020,Walters2021}. It is not clear what mechanism heats the outer layers of these white dwarfs to high temperatures, stimulating the emission. One proposal is the unipolar inductor model, in which a conducting planet induces an electrical current by orbiting inside the white dwarf's magnetosphere. The current loop connects the planet to the white dwarf's surface, which is thus heated by Ohmic dissipation \citep{Li1998,Wickramasinghe2010}. Alternatively, the emission mechanism may be intrinsic to the white dwarf, such as chromospheric activity driven by interaction between the white dwarf's magnetic field and an atmospheric convection zone \citep{GreensteinMcCarthy85,Ferrario97}. 

Interestingly, the currently known Balmer-emitting white dwarfs occupy a narrow region in the Hertzsprung--Russell (HR) diagram, clustering at temperatures $T_{\rm eff}\approx 7500\,{\rm K}$ \citep{Gansicke2020,Walters2021}.\footnote{Including J0412+7549, for which the magnetic field has not been measured yet \citep{Walters2021}.} \citet{Schreiber2021} speculated that this clustering might be related to the generation of magnetic fields by crystallization at such temperatures. 
More specifically, our Fig. \ref{fig:magnetic} suggests that the clustering of Balmer-emitting white dwarfs (which have masses of $0.6-0.8 \, {\rm M}_\odot$ according to the HR diagram; see \citealt{Gansicke2020,Walters2021}) at $T_{\rm eff}\approx 7500\,{\rm K}$ may be linked to the maximum $B_{\rm surf}(T_{\rm eff})$ attained by crystallization driven dynamos as they cool.  This maximum is reached shortly after the onset of crystallization, potentially explaining why both hotter and cooler white dwarfs lack Balmer emission lines.

Fig. \ref{fig:kawka} shows that the measured magnetic fields of the Balmer-emitters are indeed close (within an order of magnitude) to the maximal fields $B_{\rm max}$ predicted by equation \eqref{eq:super_eq} for their rotation periods. However, the measured magnetic field increases with rotation period for the three currently known Balmer-emitters, opposite to the predicted trend. Additionally, it is not clear whether the peak of $B_{\rm surf}(T_{\rm eff})$ as calculated in Fig. \ref{fig:magnetic} is sharp enough for a given $m_{\rm wd}$
and whether the crystallization temperatures are similar enough for different $m_{\rm wd}$ to explain the tight clustering around $T_{\rm eff}\approx 7500\,{\rm K}$. Moreover, the long magnetic diffusion time-scale $t_{\rm diff}$ (Section \ref{sec:diffusion}) might smear any sharp features of the $B_{\rm surf}(T_{\rm eff})$ curve. In fact, the sharp drop of $t_{\rm diff}$ at $T_{\rm eff}\approx 7500\,{\rm K}$ (for a $0.8\,{\rm M}_\odot$ white dwarf; see Fig. \ref{fig:diffusion}) suggests that the magnetic field reaches the surface at that temperature, which may provide a somewhat better explanation for the clustering of Balmer-emitters. 

\subsection{Accreting magnetic systems}

\cite{Schreiber2021Nat} proposed the crystallization-driven dynamo as the origin of strong magnetic fields in intermediate polar and polar systems, but they used the same convective turnover time-scale from \cite{Isern2017}, which is far too short. Hence, they argued that a white dwarf can only become strongly magnetized when its spin period becomes less than about a minute, such that ${\rm Ro}\lesssim 1$ and the equipartition field strength of \cite{Christensen2009} (our equation \ref{eq:equi}) can be realized. By their arguments, strong fields ($B\gtrsim$ MG) can only be produced in white dwarfs that have been spun up by accretion, explaining the far higher occurrence of magnetism in accreting systems relative to pre-CV systems (i.e. detached post-common envelope binaries).

We argue that convective turnover times are much longer, allowing moderate magnetic fields to be produced in more slowly rotating white dwarfs, but that the magnetic field increases with rotation rate according to the scaling of \cite{Augustson2019} (our equation \ref{eq:super_eq}). Hence, stronger fields should be formed in accreting systems that have been spun up, allowing the basic picture of \cite{Schreiber2021Nat} to remain valid.
A white dwarf that is spun up from $P\sim 1\,{\rm d}$ to $P\sim 1\,{\rm min}$ would have its magnetic field amplified by a factor of $\sim$30, potentially increasing its field from sub-MG levels to $\gtrsim$10 MG, transforming the system into an intermediate polar as described by \cite{Schreiber2021Nat}. A prediction of our version of the evolution is that more rapidly rotating intermediate polars should have stronger magnetic fields, on average. 

We plot the intermediate polars with measured magnetic field strengths in Fig. \ref{fig:kawka} -- all of them are consistent with our computed $B_{\rm max}$ (up to a factor of 2) and thus can be explained by a crystallization-driven dynamo if they are old enough to be crystallizing (see \citealt{Schreiber2021Nat} for a discussion). Although the period range of these intermediate polars is too narrow to test our predicted $B(P)$ relation on their own, an anti-correlation between $B$ and $P$ becomes evident when considering the relevant single white dwarfs and intermediate polars together (i.e. considering all the blue and magenta markers in the unshaded region of Fig. \ref{fig:kawka}).

What happens to the magnetic field after the white dwarf spins down to become a polar is not clear. By that point, the magnetic field must have diffused out of the dynamo region in the crystallizing core and into the stably stratified envelope above it, where it may become decoupled from the core dynamo such that it can maintain a large field strength even for a slowly rotating white dwarf (this is why we have not included polars in Fig. \ref{fig:kawka}). Unfortunately, the long-term dynamics of magnetic fields in stably stratified regions remains poorly understood, but future work may shed light on this issue.

\section{Summary}\label{sec:summary}

We have reanalysed the crystallization-driven dynamo mechanism for white dwarf magnetic field generation \citep{Isern2017}. By consistently considering the slow crystallization of the white dwarf's core, we have shown that crystallization-driven convection is much slower than previously thought, with turnover times $t_{\rm conv}\gtrsim 10^6$ s. The long turnover times imply that almost all observed white dwarfs, both single and in accreting systems, are in the fast-rotating dynamo regime, with spin periods $P\ll t_{\rm conv}$. According to dynamo theories and simulations, in this regime the magnetic energy is at least in equipartition with the kinetic energy in convective eddies, and is likely enhanced beyond equipartition by a factor of $B^2\propto t_{\rm conv}/P$ \citep{Christensen2009,Augustson2016,Augustson2019}. This rotational enhancement helps justify previously invoked strong dynamo fields and their dependence on rotation -- potentially explaining several observational puzzles, such as the abundance of magnetic CVs compared to their detached progenitors and the low effective temperatures of metal-polluted magnetic white dwarfs \citep{Schreiber2021Nat,Schreiber2021, Belloni2021}; in both cases, the white dwarfs were likely spun up by mass accretion to short $P$. 

We tracked the gradual crystallization of the white dwarf's core using the stellar evolution code \textsc{mesa} and computed $B(T_{\rm eff})$ curves. The magnetic field quickly rises to a maximum of $\sim 10^5-10^8$ G -- depending on the spin period $P$ -- at the onset of crystallization, and then gradually declines as the white dwarf cools down. When the crystallization front approaches the white dwarf's helium envelope, the dynamo shuts off and $B$ quickly drops. The range of white dwarf effective temperatures $T_{\rm eff}$ that corresponds to ongoing crystallization and dynamo activity is a function of the white dwarf's mass, with more massive white dwarfs crystallizing at higher temperatures \citep[see also][]{Schreiber2021Nat}.

We compared the maximal fields attained by crystallization-driven dynamos to observed single white dwarfs with measured magnetic fields and spin periods. About 30 per cent of the sample is consistent with a crystallization-driven dynamo up to a factor of a few in $B$, with the other white dwarfs being either too hot to be crystallizing or having magnetic fields well above our maximum, requiring a different explanation. Interestingly, the recently discovered class of single white dwarfs with Zeeman-split Balmer emission lines \citep{Gansicke2020,Walters2021} exhibits fields that are fairly close to our maximum, potentially explaining the clustering of these white dwarfs at $T_{\rm eff}\approx 7500\textrm{ K}$, where $B(T_{\rm eff})$ peaks and the magnetic diffusion time $t_{\rm diff}(T_{\rm eff})$ drops for $0.6-0.8\,{\rm M}_\odot$ white dwarfs \citep[see also][]{Schreiber2021}. 

In addition to these single white dwarfs that are candidates for hosting an active dynamo, we demonstrated that all the measured magnetic fields of intermediate polars are consistent with being generated by a crystallization-driven dynamo. When considering these intermediate polars together with low-field ($B \lesssim 10^7 \, {\rm G})$ single white dwarf candidates, the theoretical prediction $B\propto P^{-1/2}$ is roughly consistent with observations. We can also reproduce the $10^7-10^8$ G magnetic fields measured for polars, assuming they have been spun up in the past to periods of about a minute. It is not clear, however, how the magnetic field evolves after the white dwarf spins back down to synchronize with its orbit.

\section*{Acknowledgements}

We thank Wenbin Lu for interesting discussions and the anonymous reviewer for a useful report which has improved the paper. SG thanks the Heising-Simons Foundation for generous support through a 51 Pegasi b Fellowship. IC is a Sherman Fairchild Fellow at Caltech and thanks the Burke Institute at Caltech for supporting her research.

\section*{Data Availability}

The data underlying this article will be shared on reasonable request to the corresponding author.



\bibliographystyle{mnras}

\begin{thebibliography}{}
\makeatletter
\relax
\def\mn@urlcharsother{\let\do\@makeother \do\$\do\&\do\#\do\^\do\_\do\%\do\~}
\def\mn@doi{\begingroup\mn@urlcharsother \@ifnextchar [ {\mn@doi@}
  {\mn@doi@[]}}
\def\mn@doi@[#1]#2{\def\@tempa{#1}\ifx\@tempa\@empty \href
  {http://dx.doi.org/#2} {doi:#2}\else \href {http://dx.doi.org/#2} {#1}\fi
  \endgroup}
\def\mn@eprint#1#2{\mn@eprint@#1:#2::\@nil}
\def\mn@eprint@arXiv#1{\href {http://arxiv.org/abs/#1} {{\tt arXiv:#1}}}
\def\mn@eprint@dblp#1{\href {http://dblp.uni-trier.de/rec/bibtex/#1.xml}
  {dblp:#1}}
\def\mn@eprint@#1:#2:#3:#4\@nil{\def\@tempa {#1}\def\@tempb {#2}\def\@tempc
  {#3}\ifx \@tempc \@empty \let \@tempc \@tempb \let \@tempb \@tempa \fi \ifx
  \@tempb \@empty \def\@tempb {arXiv}\fi \@ifundefined
  {mn@eprint@\@tempb}{\@tempb:\@tempc}{\expandafter \expandafter \csname
  mn@eprint@\@tempb\endcsname \expandafter{\@tempc}}}

\bibitem[\protect\citeauthoryear{{Angel}, {Borra}  \& {Landstreet}}{{Angel}
  et~al.}{1981}]{Angel81}
{Angel} J.~R.~P.,  {Borra} E.~F.,   {Landstreet} J.~D.,  1981, \mn@doi [\apjs]
  {10.1086/190720}, \href
  {https://ui.adsabs.harvard.edu/abs/1981ApJS...45..457A} {45, 457}

\bibitem[\protect\citeauthoryear{{Augustson}, {Brun}  \& {Toomre}}{{Augustson}
  et~al.}{2016}]{Augustson2016}
{Augustson} K.~C.,  {Brun} A.~S.,   {Toomre} J.,  2016, \mn@doi [\apj]
  {10.3847/0004-637X/829/2/92}, \href
  {https://ui.adsabs.harvard.edu/abs/2016ApJ...829...92A} {829, 92}

\bibitem[\protect\citeauthoryear{{Augustson}, {Brun}  \& {Toomre}}{{Augustson}
  et~al.}{2019}]{Augustson2019}
{Augustson} K.~C.,  {Brun} A.~S.,   {Toomre} J.,  2019, \mn@doi [\apj]
  {10.3847/1538-4357/ab14ea}, \href
  {https://ui.adsabs.harvard.edu/abs/2019ApJ...876...83A} {876, 83}

\bibitem[\protect\citeauthoryear{{Barker}, {Dempsey}  \& {Lithwick}}{{Barker}
  et~al.}{2014}]{Barker2014}
{Barker} A.~J.,  {Dempsey} A.~M.,   {Lithwick} Y.,  2014, \mn@doi [\apj]
  {10.1088/0004-637X/791/1/13}, \href
  {https://ui.adsabs.harvard.edu/abs/2014ApJ...791...13B} {791, 13}

\bibitem[\protect\citeauthoryear{{Bauer}, {Schwab}, {Bildsten}  \&
  {Cheng}}{{Bauer} et~al.}{2020}]{Bauer2020}
{Bauer} E.~B.,  {Schwab} J.,  {Bildsten} L.,   {Cheng} S.,  2020, \mn@doi
  [\apj] {10.3847/1538-4357/abb5a5}, \href
  {https://ui.adsabs.harvard.edu/abs/2020ApJ...902...93B} {902, 93}

\bibitem[\protect\citeauthoryear{{Belloni} \& {Schreiber}}{{Belloni} \&
  {Schreiber}}{2020}]{BelloniSchreiber2020}
{Belloni} D.,  {Schreiber} M.~R.,  2020, \mn@doi [\mnras]
  {10.1093/mnras/stz3601}, \href
  {https://ui.adsabs.harvard.edu/abs/2020MNRAS.492.1523B} {492, 1523}

\bibitem[\protect\citeauthoryear{{Belloni}, {Schreiber}, {Salaris}, {Maccarone}
   \& {Zorotovic}}{{Belloni} et~al.}{2021}]{Belloni2021}
{Belloni} D.,  {Schreiber} M.~R.,  {Salaris} M.,  {Maccarone} T.~J.,
  {Zorotovic} M.,  2021, \mn@doi [\mnras] {10.1093/mnrasl/slab054}, \href
  {https://ui.adsabs.harvard.edu/abs/2021MNRAS.505L..74B} {505, L74}

\bibitem[\protect\citeauthoryear{{Braithwaite} \& {Spruit}}{{Braithwaite} \&
  {Spruit}}{2004}]{BraithwaiteSpruit2004}
{Braithwaite} J.,  {Spruit} H.~C.,  2004, \mn@doi [\nat] {10.1038/nature02934},
  \href {https://ui.adsabs.harvard.edu/abs/2004Natur.431..819B} {431, 819}

\bibitem[\protect\citeauthoryear{{Brun}, {Strugarek}, {Noraz}, {Perri},
  {Varela}, {Augustson}, {Charbonneau}  \& {Toomre}}{{Brun}
  et~al.}{2022}]{Brun2022}
{Brun} A.~S.,  {Strugarek} A.,  {Noraz} Q.,  {Perri} B.,  {Varela} J.,
  {Augustson} K.,  {Charbonneau} P.,   {Toomre} J.,  2022, \mn@doi [\apj]
  {10.3847/1538-4357/ac469b}, \href
  {https://ui.adsabs.harvard.edu/abs/2022ApJ...926...21B} {926, 21}

\bibitem[\protect\citeauthoryear{{Cantiello}, {Fuller}  \&
  {Bildsten}}{{Cantiello} et~al.}{2016}]{Cantiello2016}
{Cantiello} M.,  {Fuller} J.,   {Bildsten} L.,  2016, \mn@doi [\apj]
  {10.3847/0004-637X/824/1/14}, \href
  {https://ui.adsabs.harvard.edu/abs/2016ApJ...824...14C} {824, 14}

\bibitem[\protect\citeauthoryear{{Charbonneau} \& {MacGregor}}{{Charbonneau} \&
  {MacGregor}}{2001}]{CharbonneauMacGregor2001}
{Charbonneau} P.,  {MacGregor} K.~B.,  2001, \mn@doi [\apj] {10.1086/322417},
  \href {https://ui.adsabs.harvard.edu/abs/2001ApJ...559.1094C} {559, 1094}

\bibitem[\protect\citeauthoryear{{Christensen}}{{Christensen}}{2010}]{Christensen2010}
{Christensen} U.~R.,  2010, \mn@doi [\ssr] {10.1007/s11214-009-9553-2}, \href
  {https://ui.adsabs.harvard.edu/abs/2010SSRv..152..565C} {152, 565}

\bibitem[\protect\citeauthoryear{{Christensen} \& {Aubert}}{{Christensen} \&
  {Aubert}}{2006}]{ChristensenAubert2006}
{Christensen} U.~R.,  {Aubert} J.,  2006, \mn@doi [Geophysical Journal
  International] {10.1111/j.1365-246X.2006.03009.x}, \href
  {https://ui.adsabs.harvard.edu/abs/2006GeoJI.166...97C} {166, 97}

\bibitem[\protect\citeauthoryear{{Christensen}, {Holzwarth}  \&
  {Reiners}}{{Christensen} et~al.}{2009}]{Christensen2009}
{Christensen} U.~R.,  {Holzwarth} V.,   {Reiners} A.,  2009, \mn@doi [\nat]
  {10.1038/nature07626}, \href
  {https://ui.adsabs.harvard.edu/abs/2009Natur.457..167C} {457, 167}

\bibitem[\protect\citeauthoryear{{Cumming}}{{Cumming}}{2002}]{Cumming2002}
{Cumming} A.,  2002, \mn@doi [\mnras] {10.1046/j.1365-8711.2002.05434.x}, \href
  {https://ui.adsabs.harvard.edu/abs/2002MNRAS.333..589C} {333, 589}

\bibitem[\protect\citeauthoryear{{Davies}, {Pozzo}, {Gubbins}  \&
  {Alf{\`e}}}{{Davies} et~al.}{2015}]{Davies2015}
{Davies} C.,  {Pozzo} M.,  {Gubbins} D.,   {Alf{\`e}} D.,  2015, \mn@doi
  [Nature Geoscience] {10.1038/ngeo2492}, \href
  {https://ui.adsabs.harvard.edu/abs/2015NatGe...8..678D} {8, 678}

\bibitem[\protect\citeauthoryear{{Engdahl}, {Flinn}  \& {Mass{\'e}}}{{Engdahl}
  et~al.}{1974}]{Engdahl74}
{Engdahl} E.~R.,  {Flinn} E.~A.,   {Mass{\'e}} R.~P.,  1974, \mn@doi
  [Geophysical Journal] {10.1111/j.1365-246X.1974.tb05467.x}, \href
  {https://ui.adsabs.harvard.edu/abs/1974GeoJ...39..457E} {39, 457}

\bibitem[\protect\citeauthoryear{{Ferrario}, {Wickramasinghe}, {Liebert},
  {Schmidt}  \& {Bieging}}{{Ferrario} et~al.}{1997}]{Ferrario97}
{Ferrario} L.,  {Wickramasinghe} D.~T.,  {Liebert} J.,  {Schmidt} G.~D.,
  {Bieging} J.~H.,  1997, \mn@doi [\mnras] {10.1093/mnras/289.1.105}, \href
  {https://ui.adsabs.harvard.edu/abs/1997MNRAS.289..105F} {289, 105}

\bibitem[\protect\citeauthoryear{{Ferrario}, {de Martino}  \&
  {G{\"a}nsicke}}{{Ferrario} et~al.}{2015}]{Ferrario2015}
{Ferrario} L.,  {de Martino} D.,   {G{\"a}nsicke} B.~T.,  2015, \mn@doi [\ssr]
  {10.1007/s11214-015-0152-0}, \href
  {https://ui.adsabs.harvard.edu/abs/2015SSRv..191..111F} {191, 111}

\bibitem[\protect\citeauthoryear{{Ferrario}, {Wickramasinghe}  \&
  {Kawka}}{{Ferrario} et~al.}{2020}]{Ferrario2020}
{Ferrario} L.,  {Wickramasinghe} D.,   {Kawka} A.,  2020, \mn@doi [Advances in
  Space Research] {10.1016/j.asr.2019.11.012}, \href
  {https://ui.adsabs.harvard.edu/abs/2020AdSpR..66.1025F} {66, 1025}

\bibitem[\protect\citeauthoryear{{Finlay} \& {Amit}}{{Finlay} \&
  {Amit}}{2011}]{FinlayAmit2011}
{Finlay} C.~C.,  {Amit} H.,  2011, \mn@doi [Geophysical Journal International]
  {10.1111/j.1365-246X.2011.05032.x}, \href
  {https://ui.adsabs.harvard.edu/abs/2011GeoJI.186..175F} {186, 175}

\bibitem[\protect\citeauthoryear{{G{\"a}nsicke}, {Rodr{\'\i}guez-Gil}, {Gentile
  Fusillo}, {Inight}, {Schreiber}, {Pala}  \& {Tremblay}}{{G{\"a}nsicke}
  et~al.}{2020}]{Gansicke2020}
{G{\"a}nsicke} B.~T.,  {Rodr{\'\i}guez-Gil} P.,  {Gentile Fusillo} N.~P.,
  {Inight} K.,  {Schreiber} M.~R.,  {Pala} A.~F.,   {Tremblay} P.-E.,  2020,
  \mn@doi [\mnras] {10.1093/mnras/staa2969}, \href
  {https://ui.adsabs.harvard.edu/abs/2020MNRAS.499.2564G} {499, 2564}

\bibitem[\protect\citeauthoryear{{Garc{\'\i}a-Berro}
  et~al.,}{{Garc{\'\i}a-Berro} et~al.}{2012}]{GarciaBerro2012}
{Garc{\'\i}a-Berro} E.,  et~al., 2012, \mn@doi [\apj]
  {10.1088/0004-637X/749/1/25}, \href
  {https://ui.adsabs.harvard.edu/abs/2012ApJ...749...25G} {749, 25}

\bibitem[\protect\citeauthoryear{{Glatzmaier} \& {Roberts}}{{Glatzmaier} \&
  {Roberts}}{1997}]{GlatzmaierRoberts97}
{Glatzmaier} G.~A.,  {Roberts} P.~H.,  1997, \mn@doi [Contemporary Physics]
  {10.1080/001075197182351}, \href
  {https://ui.adsabs.harvard.edu/abs/1997ConPh..38..269G} {38, 269}

\bibitem[\protect\citeauthoryear{{Greenstein} \& {McCarthy}}{{Greenstein} \&
  {McCarthy}}{1985}]{GreensteinMcCarthy85}
{Greenstein} J.~L.,  {McCarthy} J.~K.,  1985, \mn@doi [\apj] {10.1086/162937},
  \href {https://ui.adsabs.harvard.edu/abs/1985ApJ...289..732G} {289, 732}

\bibitem[\protect\citeauthoryear{{Hollands}, {G{\"a}nsicke}  \&
  {Koester}}{{Hollands} et~al.}{2015}]{Hollands2015}
{Hollands} M.~A.,  {G{\"a}nsicke} B.~T.,   {Koester} D.,  2015, \mn@doi
  [\mnras] {10.1093/mnras/stv570}, \href
  {https://ui.adsabs.harvard.edu/abs/2015MNRAS.450..681H} {450, 681}

\bibitem[\protect\citeauthoryear{{Isern}, {Mochkovitch}, {Garc{\'\i}a-Berro}
  \& {Hernanz}}{{Isern} et~al.}{1997}]{Isern97}
{Isern} J.,  {Mochkovitch} R.,  {Garc{\'\i}a-Berro} E.,   {Hernanz} M.,  1997,
  \mn@doi [\apj] {10.1086/304425}, \href
  {https://ui.adsabs.harvard.edu/abs/1997ApJ...485..308I} {485, 308}

\bibitem[\protect\citeauthoryear{{Isern}, {Garc{\'\i}a-Berro}, {Hernanz}  \&
  {Chabrier}}{{Isern} et~al.}{2000}]{Isern2000}
{Isern} J.,  {Garc{\'\i}a-Berro} E.,  {Hernanz} M.,   {Chabrier} G.,  2000,
  \mn@doi [\apj] {10.1086/308153}, \href
  {https://ui.adsabs.harvard.edu/abs/2000ApJ...528..397I} {528, 397}

\bibitem[\protect\citeauthoryear{{Isern}, {Garc{\'\i}a-Berro}, {K{\"u}lebi}  \&
  {Lor{\'e}n-Aguilar}}{{Isern} et~al.}{2017}]{Isern2017}
{Isern} J.,  {Garc{\'\i}a-Berro} E.,  {K{\"u}lebi} B.,   {Lor{\'e}n-Aguilar}
  P.,  2017, \mn@doi [\apjl] {10.3847/2041-8213/aa5eae}, \href
  {https://ui.adsabs.harvard.edu/abs/2017ApJ...836L..28I} {836, L28}

\bibitem[\protect\citeauthoryear{{Jermyn}, {Schwab}, {Bauer}, {Timmes}  \&
  {Potekhin}}{{Jermyn} et~al.}{2021}]{Jermyn2021}
{Jermyn} A.~S.,  {Schwab} J.,  {Bauer} E.,  {Timmes} F.~X.,   {Potekhin} A.~Y.,
   2021, \mn@doi [\apj] {10.3847/1538-4357/abf48e}, \href
  {https://ui.adsabs.harvard.edu/abs/2021ApJ...913...72J} {913, 72}

\bibitem[\protect\citeauthoryear{{Kawka}}{{Kawka}}{2020}]{Kawka2020}
{Kawka} A.,  2020, \mn@doi [IAU Symposium] {10.1017/S1743921320000745}, \href
  {https://ui.adsabs.harvard.edu/abs/2020IAUS..357...60K} {357, 60}

\bibitem[\protect\citeauthoryear{{Kawka}, {Vennes}, {Ferrario}  \&
  {Paunzen}}{{Kawka} et~al.}{2019}]{Kawka2019}
{Kawka} A.,  {Vennes} S.,  {Ferrario} L.,   {Paunzen} E.,  2019, \mn@doi
  [\mnras] {10.1093/mnras/sty3048}, \href
  {https://ui.adsabs.harvard.edu/abs/2019MNRAS.482.5201K} {482, 5201}

\bibitem[\protect\citeauthoryear{{Kupka} \& {Muthsam}}{{Kupka} \&
  {Muthsam}}{2017}]{KupkaMuthsam2017}
{Kupka} F.,  {Muthsam} H.~J.,  2017, \mn@doi [Living Reviews in Computational
  Astrophysics] {10.1007/s41115-017-0001-9}, \href
  {https://ui.adsabs.harvard.edu/abs/2017LRCA....3....1K} {3, 1}

\bibitem[\protect\citeauthoryear{{Li}, {Ferrario}  \& {Wickramasinghe}}{{Li}
  et~al.}{1998}]{Li1998}
{Li} J.,  {Ferrario} L.,   {Wickramasinghe} D.,  1998, \mn@doi [\apjl]
  {10.1086/311546}, \href
  {https://ui.adsabs.harvard.edu/abs/1998ApJ...503L.151L} {503, L151}

\bibitem[\protect\citeauthoryear{{Liebert} et~al.,}{{Liebert}
  et~al.}{2005}]{Liebert2005}
{Liebert} J.,  et~al., 2005, \mn@doi [\aj] {10.1086/429639}, \href
  {https://ui.adsabs.harvard.edu/abs/2005AJ....129.2376L} {129, 2376}

\bibitem[\protect\citeauthoryear{{Liebert}, {Ferrario}, {Wickramasinghe}  \&
  {Smith}}{{Liebert} et~al.}{2015}]{Liebert2015}
{Liebert} J.,  {Ferrario} L.,  {Wickramasinghe} D.~T.,   {Smith} P.~S.,  2015,
  \mn@doi [\apj] {10.1088/0004-637X/804/2/93}, \href
  {https://ui.adsabs.harvard.edu/abs/2015ApJ...804...93L} {804, 93}

\bibitem[\protect\citeauthoryear{{Lister} \& {Buffett}}{{Lister} \&
  {Buffett}}{1995}]{ListerBuffett95}
{Lister} J.~R.,  {Buffett} B.~A.,  1995, \mn@doi [Physics of the Earth and
  Planetary Interiors] {10.1016/0031-9201(95)03042-U}, \href
  {https://ui.adsabs.harvard.edu/abs/1995PEPI...91...17L} {91, 17}

\bibitem[\protect\citeauthoryear{{Loper}}{{Loper}}{1978}]{Loper78}
{Loper} D.~E.,  1978, \mn@doi [\jgr] {10.1029/JB083iB12p05961}, \href
  {https://ui.adsabs.harvard.edu/abs/1978JGR....83.5961L} {83, 5961}

\bibitem[\protect\citeauthoryear{{MacDonald} \& {Mullan}}{{MacDonald} \&
  {Mullan}}{2004}]{MacDonaldMullan2004}
{MacDonald} J.,  {Mullan} D.~J.,  2004, \mn@doi [\mnras]
  {10.1111/j.1365-2966.2004.07394.x}, \href
  {https://ui.adsabs.harvard.edu/abs/2004MNRAS.348..702M} {348, 702}

\bibitem[\protect\citeauthoryear{{MacGregor} \& {Cassinelli}}{{MacGregor} \&
  {Cassinelli}}{2003}]{MacGregorCassinelli2003}
{MacGregor} K.~B.,  {Cassinelli} J.~P.,  2003, \mn@doi [\apj] {10.1086/346257},
  \href {https://ui.adsabs.harvard.edu/abs/2003ApJ...586..480M} {586, 480}

\bibitem[\protect\citeauthoryear{{Miller Bertolami}, {Viallet}, {Prat},
  {Barsukow}  \& {Weiss}}{{Miller Bertolami} et~al.}{2016}]{M3b2016}
{Miller Bertolami} M.~M.,  {Viallet} M.,  {Prat} V.,  {Barsukow} W.,   {Weiss}
  A.,  2016, \mn@doi [\mnras] {10.1093/mnras/stw203}, \href
  {https://ui.adsabs.harvard.edu/abs/2016MNRAS.457.4441M} {457, 4441}

\bibitem[\protect\citeauthoryear{{Mochkovitch}}{{Mochkovitch}}{1983}]{Mochkovitch83}
{Mochkovitch} R.,  1983, \aap, \href
  {https://ui.adsabs.harvard.edu/abs/1983A&A...122..212M} {122, 212}

\bibitem[\protect\citeauthoryear{{Moffatt} \& {Loper}}{{Moffatt} \&
  {Loper}}{1994}]{MoffattLoper94}
{Moffatt} H.~K.,  {Loper} D.~E.,  1994, \mn@doi [Geophysical Journal
  International] {10.1111/j.1365-246X.1994.tb03939.x}, \href
  {https://ui.adsabs.harvard.edu/abs/1994GeoJI.117..394M} {117, 394}

\bibitem[\protect\citeauthoryear{{Nandkumar} \& {Pethick}}{{Nandkumar} \&
  {Pethick}}{1984}]{NandkumarPethick84}
{Nandkumar} R.,  {Pethick} C.~J.,  1984, \mn@doi [\mnras]
  {10.1093/mnras/209.3.511}, \href
  {https://ui.adsabs.harvard.edu/abs/1984MNRAS.209..511N} {209, 511}

\bibitem[\protect\citeauthoryear{{Nordhaus}, {Wellons}, {Spiegel}, {Metzger}
  \& {Blackman}}{{Nordhaus} et~al.}{2011}]{Nordhaus2011}
{Nordhaus} J.,  {Wellons} S.,  {Spiegel} D.~S.,  {Metzger} B.~D.,   {Blackman}
  E.~G.,  2011, \mn@doi [Proceedings of the National Academy of Science]
  {10.1073/pnas.1015005108}, \href
  {https://ui.adsabs.harvard.edu/abs/2011PNAS..108.3135N} {108, 3135}

\bibitem[\protect\citeauthoryear{{Pala} et~al.,}{{Pala}
  et~al.}{2020}]{Pala2020}
{Pala} A.~F.,  et~al., 2020, \mn@doi [\mnras] {10.1093/mnras/staa764}, \href
  {https://ui.adsabs.harvard.edu/abs/2020MNRAS.494.3799P} {494, 3799}

\bibitem[\protect\citeauthoryear{{Paxton}, {Bildsten}, {Dotter}, {Herwig},
  {Lesaffre}  \& {Timmes}}{{Paxton} et~al.}{2011}]{Paxton2011}
{Paxton} B.,  {Bildsten} L.,  {Dotter} A.,  {Herwig} F.,  {Lesaffre} P.,
  {Timmes} F.,  2011, \mn@doi [\apjs] {10.1088/0067-0049/192/1/3}, \href
  {https://ui.adsabs.harvard.edu/abs/2011ApJS..192....3P} {192, 3}

\bibitem[\protect\citeauthoryear{{Paxton} et~al.,}{{Paxton}
  et~al.}{2013}]{Paxton2013}
{Paxton} B.,  et~al., 2013, \mn@doi [\apjs] {10.1088/0067-0049/208/1/4}, \href
  {https://ui.adsabs.harvard.edu/abs/2013ApJS..208....4P} {208, 4}

\bibitem[\protect\citeauthoryear{{Paxton} et~al.,}{{Paxton}
  et~al.}{2015}]{Paxton2015}
{Paxton} B.,  et~al., 2015, \mn@doi [\apjs] {10.1088/0067-0049/220/1/15}, \href
  {https://ui.adsabs.harvard.edu/abs/2015ApJS..220...15P} {220, 15}

\bibitem[\protect\citeauthoryear{{Paxton} et~al.,}{{Paxton}
  et~al.}{2018}]{Paxton2018}
{Paxton} B.,  et~al., 2018, \mn@doi [\apjs] {10.3847/1538-4365/aaa5a8}, \href
  {https://ui.adsabs.harvard.edu/abs/2018ApJS..234...34P} {234, 34}

\bibitem[\protect\citeauthoryear{{Paxton} et~al.,}{{Paxton}
  et~al.}{2019}]{Paxton2019}
{Paxton} B.,  et~al., 2019, \mn@doi [\apjs] {10.3847/1538-4365/ab2241}, \href
  {https://ui.adsabs.harvard.edu/abs/2019ApJS..243...10P} {243, 10}

\bibitem[\protect\citeauthoryear{{Potekhin} \& {Chabrier}}{{Potekhin} \&
  {Chabrier}}{2000}]{PotekhinChabrier2000}
{Potekhin} A.~Y.,  {Chabrier} G.,  2000, \mn@doi [\pre]
  {10.1103/PhysRevE.62.8554}, \href
  {https://ui.adsabs.harvard.edu/abs/2000PhRvE..62.8554P} {62, 8554}

\bibitem[\protect\citeauthoryear{{Potekhin} \& {Chabrier}}{{Potekhin} \&
  {Chabrier}}{2010}]{PotekhinChabrier2010}
{Potekhin} A.~Y.,  {Chabrier} G.,  2010, \mn@doi [Contributions to Plasma
  Physics] {10.1002/ctpp.201010017}, \href
  {https://ui.adsabs.harvard.edu/abs/2010CoPP...50...82P} {50, 82}

\bibitem[\protect\citeauthoryear{{Reding}, {Hermes}, {Vanderbosch}, {Dennihy},
  {Kaiser}, {Mace}, {Dunlap}  \& {Clemens}}{{Reding} et~al.}{2020}]{Reding2020}
{Reding} J.~S.,  {Hermes} J.~J.,  {Vanderbosch} Z.,  {Dennihy} E.,  {Kaiser}
  B.~C.,  {Mace} C.~B.,  {Dunlap} B.~H.,   {Clemens} J.~C.,  2020, \mn@doi
  [\apj] {10.3847/1538-4357/ab8239}, \href
  {https://ui.adsabs.harvard.edu/abs/2020ApJ...894...19R} {894, 19}

\bibitem[\protect\citeauthoryear{{Reg{\H{o}}s} \& {Tout}}{{Reg{\H{o}}s} \&
  {Tout}}{1995}]{RegosTout95}
{Reg{\H{o}}s} E.,  {Tout} C.~A.,  1995, \mn@doi [\mnras]
  {10.1093/mnras/273.1.146}, \href
  {https://ui.adsabs.harvard.edu/abs/1995MNRAS.273..146R} {273, 146}

\bibitem[\protect\citeauthoryear{{Reiners} \& {Christensen}}{{Reiners} \&
  {Christensen}}{2010}]{ReinersChristensen2010}
{Reiners} A.,  {Christensen} U.~R.,  2010, \mn@doi [\aap]
  {10.1051/0004-6361/201014251}, \href
  {https://ui.adsabs.harvard.edu/abs/2010A&A...522A..13R} {522, A13}

\bibitem[\protect\citeauthoryear{{Salaris}, {Dom{\'\i}nguez},
  {Garc{\'\i}a-Berro}, {Hernanz}, {Isern}  \& {Mochkovitch}}{{Salaris}
  et~al.}{1997}]{Salaris97}
{Salaris} M.,  {Dom{\'\i}nguez} I.,  {Garc{\'\i}a-Berro} E.,  {Hernanz} M.,
  {Isern} J.,   {Mochkovitch} R.,  1997, \mn@doi [\apj] {10.1086/304483}, \href
  {https://ui.adsabs.harvard.edu/abs/1997ApJ...486..413S} {486, 413}

\bibitem[\protect\citeauthoryear{{Salaris}, {Cassisi}, {Pietrinferni},
  {Kowalski}  \& {Isern}}{{Salaris} et~al.}{2010}]{Salaris2010}
{Salaris} M.,  {Cassisi} S.,  {Pietrinferni} A.,  {Kowalski} P.~M.,   {Isern}
  J.,  2010, \mn@doi [\apj] {10.1088/0004-637X/716/2/1241}, \href
  {https://ui.adsabs.harvard.edu/abs/2010ApJ...716.1241S} {716, 1241}

\bibitem[\protect\citeauthoryear{{Schaeffer}, {Jault}, {Nataf}  \&
  {Fournier}}{{Schaeffer} et~al.}{2017}]{Schaeffer2017}
{Schaeffer} N.,  {Jault} D.,  {Nataf} H.~C.,   {Fournier} A.,  2017, \mn@doi
  [Geophysical Journal International] {10.1093/gji/ggx265}, \href
  {https://ui.adsabs.harvard.edu/abs/2017GeoJI.211....1S} {211, 1}

\bibitem[\protect\citeauthoryear{{Schreiber}, {Belloni}, {G{\"a}nsicke},
  {Parsons}  \& {Zorotovic}}{{Schreiber} et~al.}{2021a}]{Schreiber2021Nat}
{Schreiber} M.~R.,  {Belloni} D.,  {G{\"a}nsicke} B.~T.,  {Parsons} S.~G.,
  {Zorotovic} M.,  2021a, \mn@doi [Nature Astronomy]
  {10.1038/s41550-021-01346-8}, \href
  {https://ui.adsabs.harvard.edu/abs/2021NatAs...5..648S} {5, 648}

\bibitem[\protect\citeauthoryear{{Schreiber}, {Belloni}, {G{\"a}nsicke}  \&
  {Parsons}}{{Schreiber} et~al.}{2021b}]{Schreiber2021}
{Schreiber} M.~R.,  {Belloni} D.,  {G{\"a}nsicke} B.~T.,   {Parsons} S.~G.,
  2021b, \mn@doi [\mnras] {10.1093/mnrasl/slab069}, \href
  {https://ui.adsabs.harvard.edu/abs/2021MNRAS.506L..29S} {506, L29}

\bibitem[\protect\citeauthoryear{{Spitzer}}{{Spitzer}}{1962}]{Spitzer62}
{Spitzer} L.,  1962, {Physics of Fully Ionized Gases}.
Interscience, New York

\bibitem[\protect\citeauthoryear{{Stevenson}}{{Stevenson}}{1979}]{Stevenson1979}
{Stevenson} D.~J.,  1979, \mn@doi [Geophysical and Astrophysical Fluid
  Dynamics] {10.1080/03091927908242681}, \href
  {https://ui.adsabs.harvard.edu/abs/1979GApFD..12..139S} {12, 139}

\bibitem[\protect\citeauthoryear{{Stevenson}}{{Stevenson}}{1980}]{Stevenson1980}
{Stevenson} D.~J.,  1980, Journal de Physique, \href
  {https://ui.adsabs.harvard.edu/abs/1980JPhys..41C..61S} {41, C2, 61}

\bibitem[\protect\citeauthoryear{{Stevenson}, {Spohn}  \&
  {Schubert}}{{Stevenson} et~al.}{1983}]{Stevenson83}
{Stevenson} D.~J.,  {Spohn} T.,   {Schubert} G.,  1983, \mn@doi [\icarus]
  {10.1016/0019-1035(83)90241-5}, \href
  {https://ui.adsabs.harvard.edu/abs/1983Icar...54..466S} {54, 466}

\bibitem[\protect\citeauthoryear{{Straniero}, {Dom{\'\i}nguez}, {Imbriani}  \&
  {Piersanti}}{{Straniero} et~al.}{2003}]{Straniero2003}
{Straniero} O.,  {Dom{\'\i}nguez} I.,  {Imbriani} G.,   {Piersanti} L.,  2003,
  \mn@doi [\apj] {10.1086/345427}, \href
  {https://ui.adsabs.harvard.edu/abs/2003ApJ...583..878S} {583, 878}

\bibitem[\protect\citeauthoryear{{Tout}, {Wickramasinghe}  \&
  {Ferrario}}{{Tout} et~al.}{2004}]{Tout2004}
{Tout} C.~A.,  {Wickramasinghe} D.~T.,   {Ferrario} L.,  2004, \mn@doi [\mnras]
  {10.1111/j.1365-2966.2004.08482.x}, \href
  {https://ui.adsabs.harvard.edu/abs/2004MNRAS.355L..13T} {355, L13}

\bibitem[\protect\citeauthoryear{{Tout}, {Wickramasinghe}, {Liebert},
  {Ferrario}  \& {Pringle}}{{Tout} et~al.}{2008}]{Tout2008}
{Tout} C.~A.,  {Wickramasinghe} D.~T.,  {Liebert} J.,  {Ferrario} L.,
  {Pringle} J.~E.,  2008, \mn@doi [\mnras] {10.1111/j.1365-2966.2008.13291.x},
  \href {https://ui.adsabs.harvard.edu/abs/2008MNRAS.387..897T} {387, 897}

\bibitem[\protect\citeauthoryear{{Walters} et~al.,}{{Walters}
  et~al.}{2021}]{Walters2021}
{Walters} N.,  et~al., 2021, \mn@doi [\mnras] {10.1093/mnras/stab617}, \href
  {https://ui.adsabs.harvard.edu/abs/2021MNRAS.503.3743W} {503, 3743}

\bibitem[\protect\citeauthoryear{{Wendell}, {van Horn}  \& {Sargent}}{{Wendell}
  et~al.}{1987}]{Wendell1987}
{Wendell} C.~E.,  {van Horn} H.~M.,   {Sargent} D.,  1987, \mn@doi [\apj]
  {10.1086/164968}, \href
  {https://ui.adsabs.harvard.edu/abs/1987ApJ...313..284W} {313, 284}

\bibitem[\protect\citeauthoryear{{Wickramasinghe} \&
  {Ferrario}}{{Wickramasinghe} \&
  {Ferrario}}{2005}]{WickramasingheFerrario2005}
{Wickramasinghe} D.~T.,  {Ferrario} L.,  2005, \mn@doi [\mnras]
  {10.1111/j.1365-2966.2004.08603.x}, \href
  {https://ui.adsabs.harvard.edu/abs/2005MNRAS.356.1576W} {356, 1576}

\bibitem[\protect\citeauthoryear{{Wickramasinghe}, {Farihi}, {Tout}, {Ferrario}
   \& {Stancliffe}}{{Wickramasinghe} et~al.}{2010}]{Wickramasinghe2010}
{Wickramasinghe} D.~T.,  {Farihi} J.,  {Tout} C.~A.,  {Ferrario} L.,
  {Stancliffe} R.~J.,  2010, \mn@doi [\mnras]
  {10.1111/j.1365-2966.2010.16417.x}, \href
  {https://ui.adsabs.harvard.edu/abs/2010MNRAS.404.1984W} {404, 1984}

\makeatother
\end{thebibliography}
\input{crystal.bbl}



\appendix

\section{Outer edge of convection}\label{sec:r_out}

\begin{figure}
\includegraphics[width=\columnwidth]{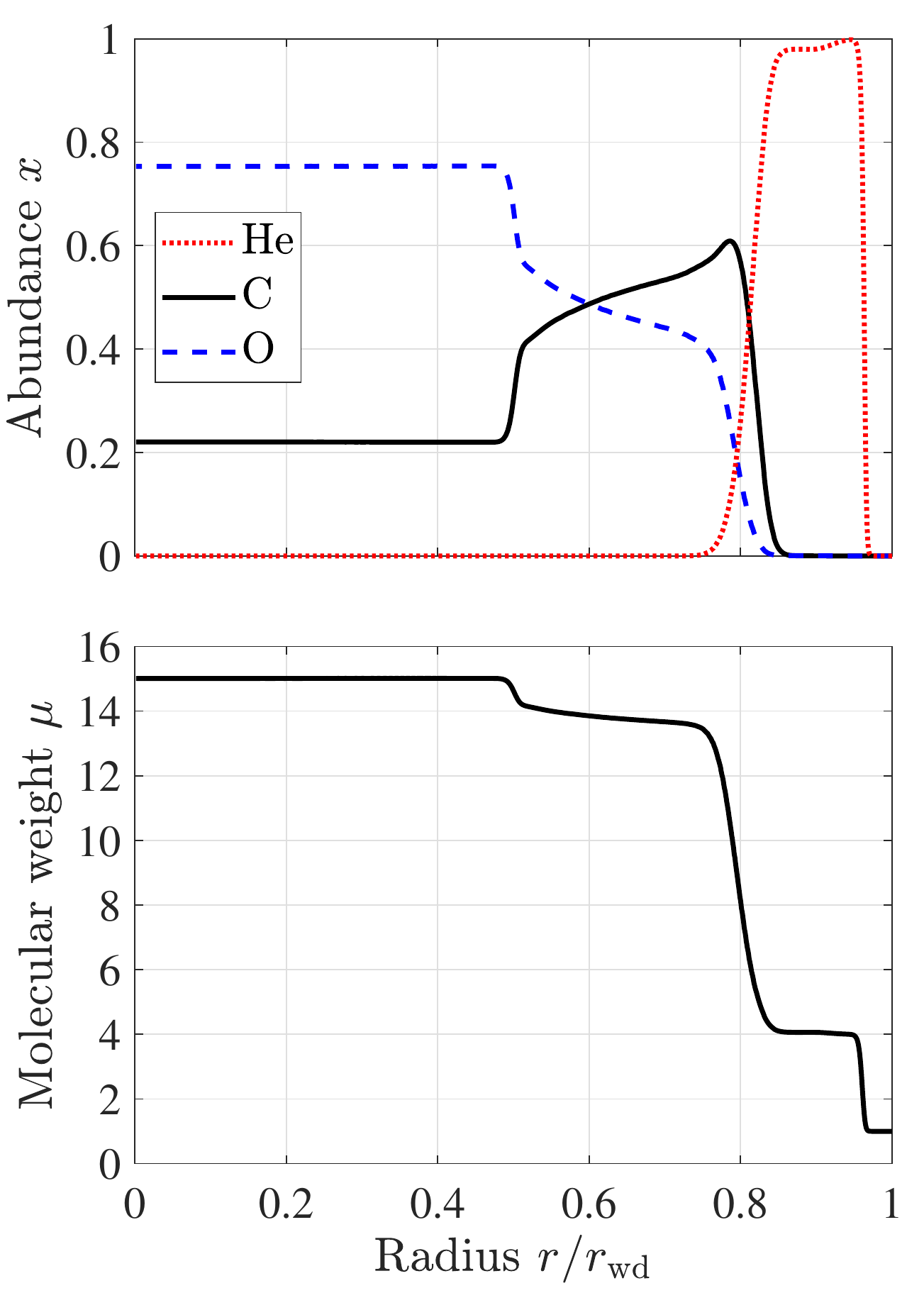}
\caption{Helium, carbon, and oxygen abundances inside a $0.6\,{\rm M}_\odot$ white dwarf at the onset of crystallization (top panel), as well as the mean molecular weight (bottom panel). During most of crystallization, convection stops at a steep $\mu$ gradient: $r_{\rm out}\approx 0.5\,r_{\rm wd}$ at early times (high $T_{\rm eff}$), corresponding to the jump in the C to O ratio, and $r_{\rm out}\approx 0.8\,r_{\rm wd}$ at late times (low $T_{\rm eff}$), corresponding to the inner edge of the helium layer; see also Fig. \ref{fig:scales}. The same behaviour is exhibited (qualitatively) by $0.8\,{\rm M}_\odot$ white dwarfs as well.}
\label{fig:fracs}
\end{figure}

Fig. \ref{fig:fracs} shows the element distribution and the mean molecular weight $\mu$ inside a white dwarf at the onset of crystallization. As the core of the white dwarf, up to a radius $r$ and a mass $m$, crystallizes into an oxygen rich solid, the liquid right above it is enriched in carbon. Crystallization therefore lowers $\mu$ on top of the core compared to the ambient CO liquid further above -- driving convection. Convection reaches an outer radius $r_{\rm out}$, where the ambient molecular weight $\mu_{\rm out}\equiv\mu(r_{\rm out})$ drops below that of the convection zone $\mu_{\rm conv}$ -- stabilizing the liquid ($\mu$ is assumed to be uniform in the convection zone due to mixing by eddies).

At each step of the white dwarf's evolution (as the crystallized $r$ and $m$ grow), we calculate $r_{\rm out}$ consistently by solving
\begin{equation}\label{eq:mu}
    \mu_{\rm conv}(r_{\rm out})=\mu_{\rm out}(r_{\rm out}).
\end{equation}
The mean molecular weight in the convective zone $\mu_{\rm conv}$ is calculated by assuming that the carbon mass between $r$ and $r_{\rm out}$ is enriched by $m\Delta x$ whereas the oxygen mass is depleted by a similar amount; the enrichment in crystallization is $\Delta x\approx0.2$ \citep{Isern2000}. $\mu_{\rm out}$ is simply the unperturbed $\mu(r_{\rm out}$). We solve equation \eqref{eq:mu} by considering progressively larger $r_{\rm out}>r$ until the stabilizing condition is satisfied. 

Our resulting $r_{\rm out}$ (Fig. \ref{fig:scales}) behaves similarly to \citet{Isern2017}. It follows the steep $\mu$ gradient produced by the jump\footnote{The inner homogeneous region seen in Fig. \ref{fig:fracs} is a relic of convective mixing during core helium burning in the white dwarf's progenitor star \citep[e.g.][]{Salaris97,Straniero2003}.} in the C to O ratio at early times (high $T_{\rm eff}$), and the inner edge of the helium envelope at late times (low $T_{\rm eff}$); see Fig. \ref{fig:fracs}.    


\bsp	
\label{lastpage}
\end{document}